\documentclass[twocolumn]{aastex62}
\usepackage{color,soul}
\usepackage{comment}
\usepackage{amsmath}
\usepackage{tablefootnote}

\graphicspath{{./}{finalfigures/}{figures/}}

\submitjournal{ApJ}

\shorttitle{The Star Formation History and Nucleosynthesis of Sculptor dSph}
\shortauthors{de los Reyes et al.}

\begin{document}

\title{Simultaneous Constraints on the Star Formation History and Nucleosynthesis of Sculptor dSph}

\correspondingauthor{Mithi A. C. de los Reyes}
\email{mdelosre@caltech.edu}

\author[0000-0002-4739-046X]{Mithi A. C. de los Reyes}
\affiliation{California Institute of Technology, 1200 E. California Blvd., MC 249-17, Pasadena, CA 91125, USA}

\author[0000-0001-6196-5162]{Evan N. Kirby}
\affiliation{California Institute of Technology, 1200 E. California Blvd., MC 249-17, Pasadena, CA 91125, USA}
\affiliation{Department of Physics, University of Notre Dame, Notre Dame, IN 46556, USA}

\author[0000-0002-4863-8842]{Alexander P. Ji}
\affiliation{Observatories of the Carnegie Institution for Science, 813 Santa Barbara St., Pasadena, CA 91101, USA}
\affiliation{Department of Astronomy \& Astrophysics, University of Chicago, 5640 S. Ellis Avenue, Chicago, IL 60637, USA}
\affiliation{Kavli Institute for Cosmological Physics, University of Chicago, Chicago, IL 60637, USA}

\author[0000-0001-5595-757X]{Evan H. Nu\~{n}ez}
\affiliation{California Institute of Technology, 1200 E. California Blvd., MC 249-17, Pasadena, CA 91125, USA}

\begin{abstract}
We demonstrate that using up to seven stellar abundance ratios can place observational constraints on the star formation histories (SFHs) of Local Group dSphs, using the Sculptor dSph as a test case.
We use a one-zone chemical evolution model to fit the overall abundance patterns of $\alpha$ elements (which probe the core-collapse supernovae that occur shortly after star formation), $s$-process elements (which probe AGB nucleosynthesis at intermediate delay times), and iron-peak elements (which probe delayed Type Ia supernovae).
Our best-fit model indicates that Sculptor dSph has an ancient SFH, consistent with previous estimates from deep photometry.
However, we derive a total star formation duration of $\sim0.9$~Gyr, which is shorter than photometrically derived SFHs.
We explore the effect of various model assumptions on our measurement and find that modifications to these assumptions still produce relatively short SFHs of duration $\lesssim1.4$~Gyr.
Our model is also able to compare sets of predicted nucleosynthetic yields for supernovae and AGB stars, and can provide insight into the nucleosynthesis of individual elements in Sculptor dSph.
We find that observed [Mn/Fe] and [Ni/Fe] trends are most consistent with sub-$M_{\mathrm{Ch}}$ Type Ia supernova models, and that a combination of ``prompt'' (delay times similar to core-collapse supernovae) and ``delayed'' (minimum delay times $\gtrsim50$~Myr) $r$-process events may be required to reproduce observed [Ba/Mg] and [Eu/Mg] trends.\\
\end{abstract}

\section{Introduction} 
\label{sec:intro}

Inside a star-forming galaxy, baryonic matter is constantly cycling between two phases of matter: stars and the interstellar medium (ISM).
The ISM, which is predominantly gas, contains the raw material that forms stars; stars produce heavy elements throughout their lifetimes, then release them back into the ISM when they die.
This cycle is not closed---stars can also produce outflows that remove gas from a galaxy \citep[][]{Mathews1971,Larson1974}, and inflows of gas can add metal-poor material to a galaxy \citep[][]{Larson1972,Dekel2009}.
It is also not the only physical process driving galaxy evolution---environmental effects such as galaxy mergers \citep[e.g.,][]{TinsleyLarson1979} and ram pressure stripping \citep[e.g.,][]{Lin1983}, and activity driven by supermassive black holes \citep[][]{Fabian2012} may have dramatic impacts on a galaxy's history. 
Yet the star formation cycle underpins the evolution of every galaxy, and measuring how star formation rates change over time---galaxy \emph{star formation histories} (SFHs)---is therefore critical to interpreting galaxy evolution.

There are a number of methods for measuring galaxy SFHs. 
First, galaxy SFHs may be estimated by fitting integrated spectral energy distributions (SEDs) with models derived from stellar population synthesis \citep[for reviews, see][]{Walcher2011,Conroy2013}.
The SEDs may be composed of broadband photometry \citep[e.g.,][]{Smith2015} or a continuous spectrum of the integrated light of the galaxy's stars and ionized gas \citep[e.g.,][]{Magris2015}.
Although this method is useful for obtaining SFHs of distant, unresolved galaxies, it depends strongly on prior assumptions about the model SFHs \citep{Carnall2019,Leja2019}, as well as the stellar initial mass function (IMF) \citep[][among others]{Conroy2012}.

For nearby galaxies that can be resolved photometrically, SFHs can be more robustly derived by fitting isochrones, which depend on stellar ages and metallicities, to observed color-magnitude diagrams (CMDs).
This method has been used to measure SFHs of many of the galaxies in the Local Group \citep[e.g.,][]{Weisz2014}.
However, this method struggles to obtain precise SFHs for galaxies with predominantly old or metal-poor populations, because isochrones are roughly logarithmically spaced in age and metallicity.  
For example, the difference between otherwise identical isochrones at 11~Gyr compared to 13~Gyr (i.e., 10-20\% accuracy and precision) is only a few hundredths of a magnitude in commonly used broadband filters---smaller than the discrepancies between different sets of isochrones modeled with the same parameters.
Metal-poor isochrones ($\mathrm{[Fe/H]} \lesssim -2$) similarly bunch together in the CMD\@.
One mitigation strategy is to obtain spectroscopic abundances for individual stars.
These abundances can be used to fix the metallicities (and sometimes detailed abundance ratios) of the isochrones so that the problem is reduced to measuring age alone rather than age and metallicity simultaneously. 
This technique works either by measuring the ages of individual stars with known spectroscopic metallicities \citep{Kirby2017} or by using a spectroscopic metallicity distribution of a representative subset of the stars being fit in the CMD \citep[e.g.,][]{Brown2014}.

\begin{figure}[t!]
    \centering
    \epsscale{1.15}
    \plotone{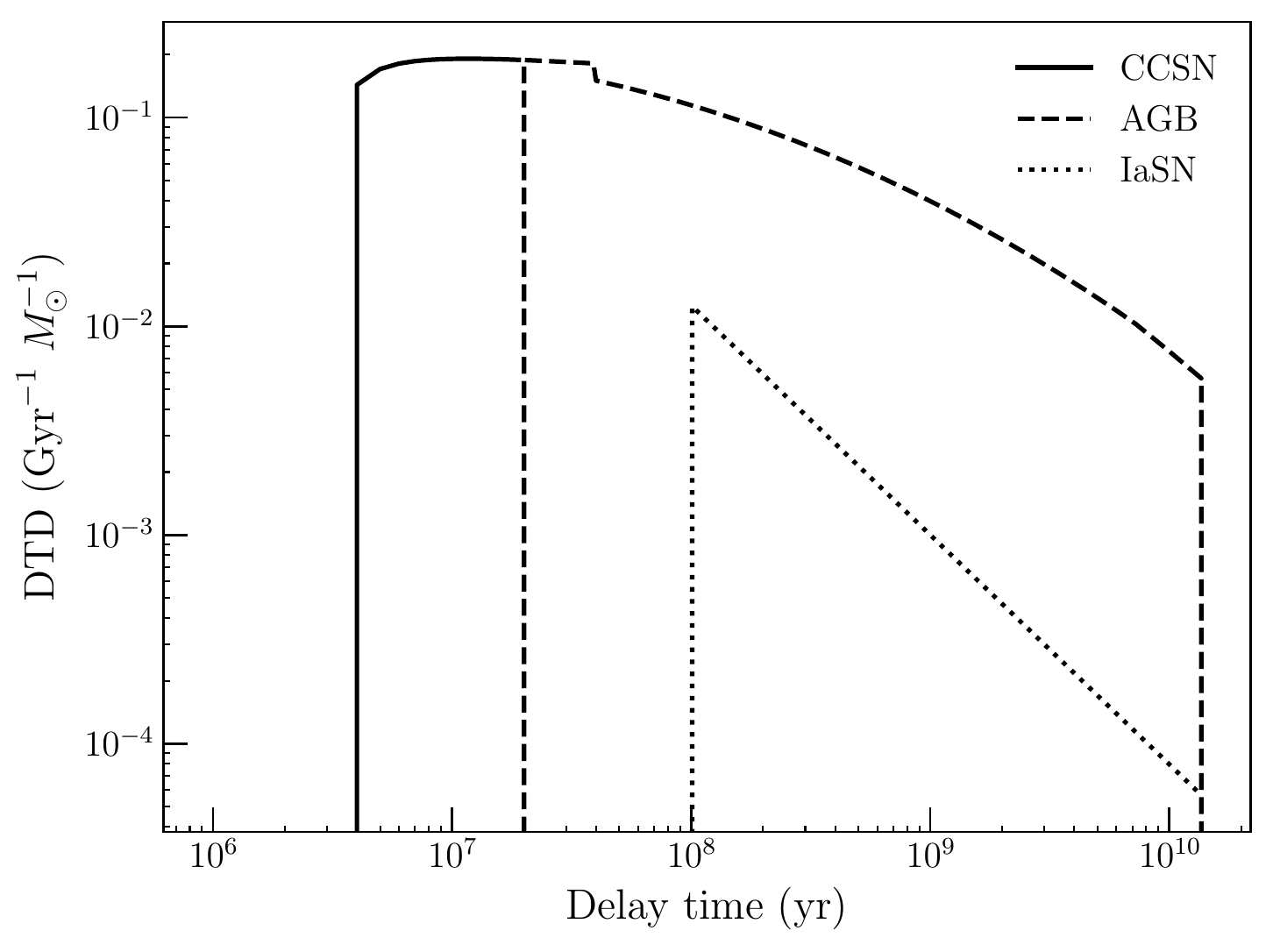}
    \caption{Delay-time distributions (DTDs) of core-collapse supernovae, AGB stars, and Type Ia supernovae, illustrating the different timescales that these events probe. The slight discontinuity in the AGB DTD arises from the different equations used to describe stellar lifetimes for stars with masses above and below $6.6~M_{\odot}$ (Equations~\ref{eq:lifetime_himass} and \ref{eq:lifetime_lomass}).}
    \label{fig:dtd}
\end{figure}

\begin{figure}[t!]
    \centering
    \epsscale{1.1}
    \plotone{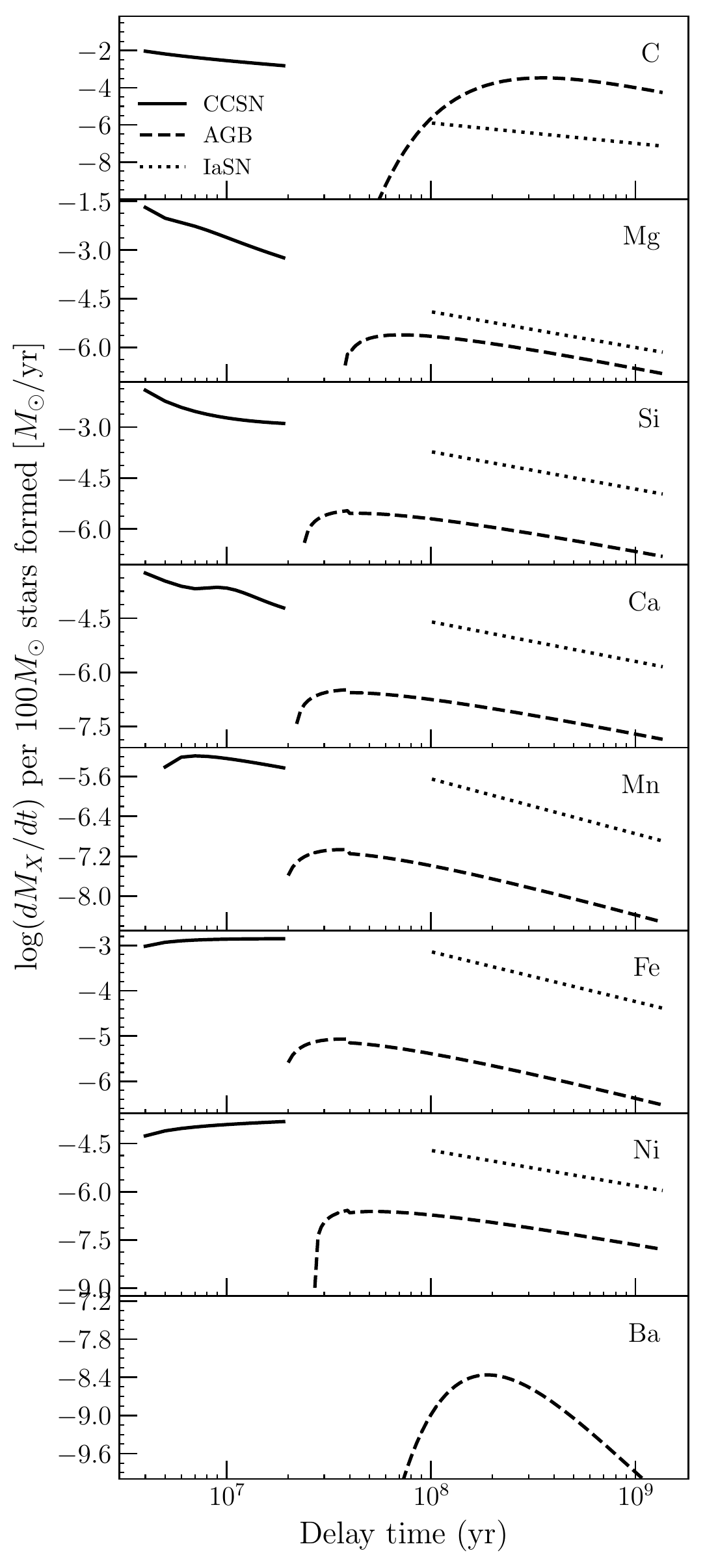}
    \caption{The ``impulse response'' of elemental yields to an instantaneous $100M_{\odot}$ burst of star formation---that is, the rate at which individual elements are produced as a function of delay time after $100M_{\odot}$ of stars are formed at $t=0$. The yields from different nucleosynthetic events (core-collapse SNe, AGB stars, and Type Ia SNe) are plotted as a function of delay time after star formation. We assume a Type Ia delay-time distribution, stellar lifetimes, a stellar IMF, and parameterized stellar yields as described in Section~\ref{sec:methods}. For simplicity, we also assume here that all nucleosynthetic events have a single metallicity of $Z=0.001Z_{\odot}$.}
    \label{fig:impulse}
\end{figure}

Galactic chemical evolution (GCE) models represent a complementary approach to measuring SFHs.
By modeling nucleosynthetic events that occur throughout a galaxy's history, GCE models predict stellar abundance trends that can then be compared with abundances measured from observed stellar spectra.
Stellar abundance trends are sensitive to the SFH because, as shown in Figure~\ref{fig:dtd}, different types of nucleosynthetic events occur at different delay times after a burst of star formation \citep[e.g.,][]{Tinsley1979,Gilmore1991}.
We illustrate this in Figure~\ref{fig:impulse}, which shows the masses of different elements produced in response to an instantaneous burst of star formation.
For example, core-collapse supernovae (CCSNe) explode on short timescales of tens of millions of years after a burst of star formation (solid lines in Figures~\ref{fig:dtd} and \ref{fig:impulse}).
The $\alpha$-elements (e.g., Mg, Ca, Si, Ti) that are predominantly produced in CCSNe are therefore indicators of the chemical enrichment that most immediately follows an episode of star formation.
Iron-peak elements, on the other hand, are largely produced by Type Ia SNe.
Since nucleosynthetic yields from Type Ia SNe dominate at late times \citep[$\gtrsim100$~Myr after star formation; see, e.g.,][]{Maoz2012}, iron-peak elements trace the most delayed times in a galaxy's SFH (dotted lines in Figures~\ref{fig:dtd} and \ref{fig:impulse}). 

\citet{Pagel2006} and \citet{Matteucci2012} describe the ingredients of a GCE model. 
The first such models can be traced back to \citet{Tinsley1968}; since then, a veritable cornucopia of chemical evolution models have been developed, particularly for Local Group dwarf galaxies \citep[for a more detailed review of chemical evolution models of dwarf galaxies before the last decade, we refer the reader to][] {Tolstoy2009}.
Local Group dSphs are, in many ways, ideal systems to study using chemical evolution models.
Many Local Group dSphs have simple SFHs---typically one or a few bursts of star formation, followed by relatively low levels of star formation \citep[e.g.,][]{Weisz2014}.
Most dwarf galaxies are also well-mixed \citep{Escala2018} and can be reasonably well described by one-zone GCE models.
Finally, recent stellar spectroscopic surveys \citep[e.g.,][]{Kirby2010,Majewski2017,Hill2019} have obtained a large number of dSph stellar spectra, and upcoming surveys and instruments---Subaru PFS \citep[][]{PFS2018}, VLT/MOONS \citep[][]{MOONS2018}, and MSE \citep[][]{MSE2016}, among others---promise to obtain many more.

Several of these dwarf galaxy models are relatively simple one-zone chemical evolution models that have attempted to match elemental abundance patterns in Local Group dSphs \citep[e.g.,][]{Carigi2002,Lanfranchi2003,Lanfranchi2004,Fenner2006,Kirby2011,Vincenzo2014,Homma2015,Kobayashi2015,Ural2015,Cote2017} or disrupted dwarf galaxies \citep[e.g., the Gaia-Enceladus-Sausage;][]{Vincenzo2019}.
Many of these simple models have been extremely successful; in fact, \citet{Vincenzo2016} showed that the outputs from such a one-zone model can reproduce many of the observed \emph{photometric} features on a dSph's CMD.
More recently, some one-zone chemical evolution models have attempted to match \emph{isotopic} abundance patterns \citep[][]{Pandey2021}.
Other models use input parameters from semianalytic models of galaxy formation \citep[e.g.,][]{Calura2009,Romano2013} or are chemodynamical (and hydrodynamical), tracking both the kinematics and abundances in dwarf galaxies \citep[e.g.,][]{Recchi2001,Kawata2006,Marcolini2006,Revaz2009,Revaz2016,Revaz2012,Hirai2017,Escala2018}.

However, many chemical evolution models of Local Group dSphs require assumptions about the SFH\@.
One-zone models in particular typically use the SFHs determined from photometric studies as inputs.
For example, the majority of previous one-zone models of Sculptor dSph \citep[][]{Lanfranchi2003,Lanfranchi2004,Fenner2006,Vincenzo2014,Homma2015,Cote2017,Pandey2021} have assumed SFHs from CMD studies \citep{Dolphin2002,deBoer2012}.
The old, metal-poor stellar populations that dominate dSphs make it difficult to obtain photometrically derived SFHs with fine time resolution.
In this paper, we instead use a one-zone GCE model to independently derive the SFHs of dSphs in the Local Group, expanding upon previous work by, e.g., \citet{Kirby2011} and \citet{Vincenzo2016}.

This work demonstrates how simultaneously fitting a wide variety of stellar abundances can place useful constraints on the SFHs of Local Group dSphs, using Sculptor dSph as a test case.
We use a one-zone chemical evolution model that, like previous dSph models, is able to fit the overall abundance patterns of $\alpha$ elements and iron-peak elements in these galaxies.
Unlike most previous one-zone models, we also use our model to fit the abundances of carbon and barium, elements predominantly produced by asymptotic giant branch (AGB) stars.
Medium-resolution spectroscopy has enabled large homogeneous catalogs of abundances of C and Ba in dSphs.
These elements probe \emph{intermediate} delay times (dashed lines in Figures~\ref{fig:dtd} and \ref{fig:impulse}) after a burst of star formation and are therefore crucial to constraining a galaxy's full SFH.

Additionally, like other GCE models, our model can provide insights into chemical evolution in Sculptor dSph.
A major hurdle for GCE models is disentangling the contributions from multiple nucleosynthetic channels; as shown in Figure~\ref{fig:impulse}, a single element may be produced by multiple nucleosynthetic sources.
This is further complicated by significant uncertainties in additional nucleosynthetic processes, such as the $r$-process and $i$-process.
We aim to build on the results of previous analyses of abundance trends \citep[e.g.,][]{Kirby2011,Kirby2019,Duggan2018,Hill2019,Skuladottir2019,delosReyes2020} and one-zone GCE models \citep[e.g.,][]{Cote2017} of Sculptor dSph, not only by simultaneously fitting several types of elements---particularly elements produced by AGB stars, which have not typically been included in GCE models of dSphs---but also by varying the input yields from supernovae and AGB stars.
This will allow us to directly compare theoretical yield sets.
We can also compare our model to the observed abundance trends of elements that were not used to fit our model but that may be sensitive to particular nucleosynthetic channels: for example, manganese and nickel are sensitive to the density of Type Ia supernova progenitors \citep[e.g.,][]{Seitenzahl2013}, while barium and europium are produced by the $r$-process.

The structure of this paper is as follows.
We describe the observed chemical abundances and the simple GCE model used to fit these abundances in Section~\ref{sec:methods} before presenting the measured SFH of Sculptor dSph in Section~\ref{sec:results}.
In Section~\ref{sec:discussion} we compare our measured SFH to previous literature measurements and discuss the effects of our model assumptions.
Our simple GCE model also allows us to probe the nucleosynthesis of different individual elements, and we discuss these additional implications in Section~\ref{sec:nucleosynthesis}.
We summarize our conclusions in Section~\ref{sec:conclusion}.
Throughout this paper, we assume a flat $\Lambda$CDM cosmology with Planck 2018 parameters \citep[$H_{0}=67.4~\mathrm{km}~\mathrm{s}^{-1}\mathrm{Mpc}^{-1}$, $\Omega_{m}=0.315$;][]{ Planck2018}. 

\section{Methods} 
\label{sec:methods}

\subsection{Abundance Measurements}
\label{sec:observations}

In this work, we primarily use literature abundances derived from medium-resolution spectroscopy with the DEep Imaging Multi-Object Spectrograph \citep[DEIMOS;][]{Faber2003} on the Keck II telescope.
A number of previous works have obtained spectra of red giant branch stars in several globular clusters and classical dSphs.
We compile several abundance ratios\footnote{Throughout this paper, we use bracket abundances referenced to solar (e.g., [Fe/H] = $\log_{10}(n_{\mathrm{Fe}}/n_{\mathrm{H}})_{\ast}-\log_{10}(n_{\mathrm{Fe}}/n_{\mathrm{H}})_{\odot}$), where $n_{\mathrm{X}}$ is the atomic number density of X. Solar abundances are adopted from \citet{Asplund2009}.} from these catalogs: [Fe/H] and [$\alpha$/Fe] ([Mg/Fe], [Si/Fe], and [Ca/Fe]) abundances from \citet{Kirby2010}, [C/Fe] from \citet{Kirby2015}, [Ni/Fe] from \citet{Kirby2018}, [Mn/Fe] from \citet{delosReyes2020}, and [Ba/Fe] from \citet{Duggan2018}.
We also use supplemental data from the DART survey \citep{Tolstoy2006}, which used ESO VLT/FLAMES to obtain high-resolution ($R\gtrsim20,000$) spectra of RGB stars in dSphs.
The DART abundance ratios for Sculptor dSph are presented in \citet{North2012}, who measured [Mn/Fe], and \citet{Hill2019}, who measured all the other abundances used in this work.

In particular, we use the DART data to modify the [Ba/Fe] abundances for our analysis.
Although the majority of barium is produced in the $s$-process in AGB stars \citep[see, e.g., Table 10 of][]{Simmerer2004}, $r$-process nucleosynthesis also contributes to the production of barium---particularly at low metallicities---and the sites and yields of $r$-process nucleosynthesis are poorly constrained \citep[][]{Cowan2021}.
For this reason, we opt not to include the $r$-process in our GCE model.
To accommodate this choice, we remove the $r$-process contributions to the barium yields by using measurements of europium (Eu), which is almost entirely produced by the $r$-process \citep[][]{Simmerer2004}.
We use the available Eu measurements from the DART survey \citep[][]{Hill2019} to compute [Ba/Eu], which is an indicator of the ratio of $s$-process to $r$-process contributions.
We follow the procedure outlined in \citet{Duggan2018} to convert [Ba/Eu] to the fraction of barium produced from the $r$-process:
\begin{equation}
    f_{r} = \frac{\frac{N_{s}(\mathrm{Eu})}{N_{s}(\mathrm{Ba})}-10^{\mathrm{[Ba/Eu]}_{\odot}-\mathrm{[Ba/Eu]}}}{\frac{N_{s}(\mathrm{Eu})}{N_{s}(\mathrm{Ba})} - \frac{N_{r}(\mathrm{Eu})}{N_{r}(\mathrm{Ba})}}
    \label{eq:bacorr}
\end{equation}
where $N_{s}(\mathrm{X})$ and $N_{r}(\mathrm{X})$ are the solar $s$-process and $r$-process number abundances of element X, obtained from Table 10 of \citet{Simmerer2004}.

\begin{figure}[t!]
    \centering
    \epsscale{1.15}
    \plotone{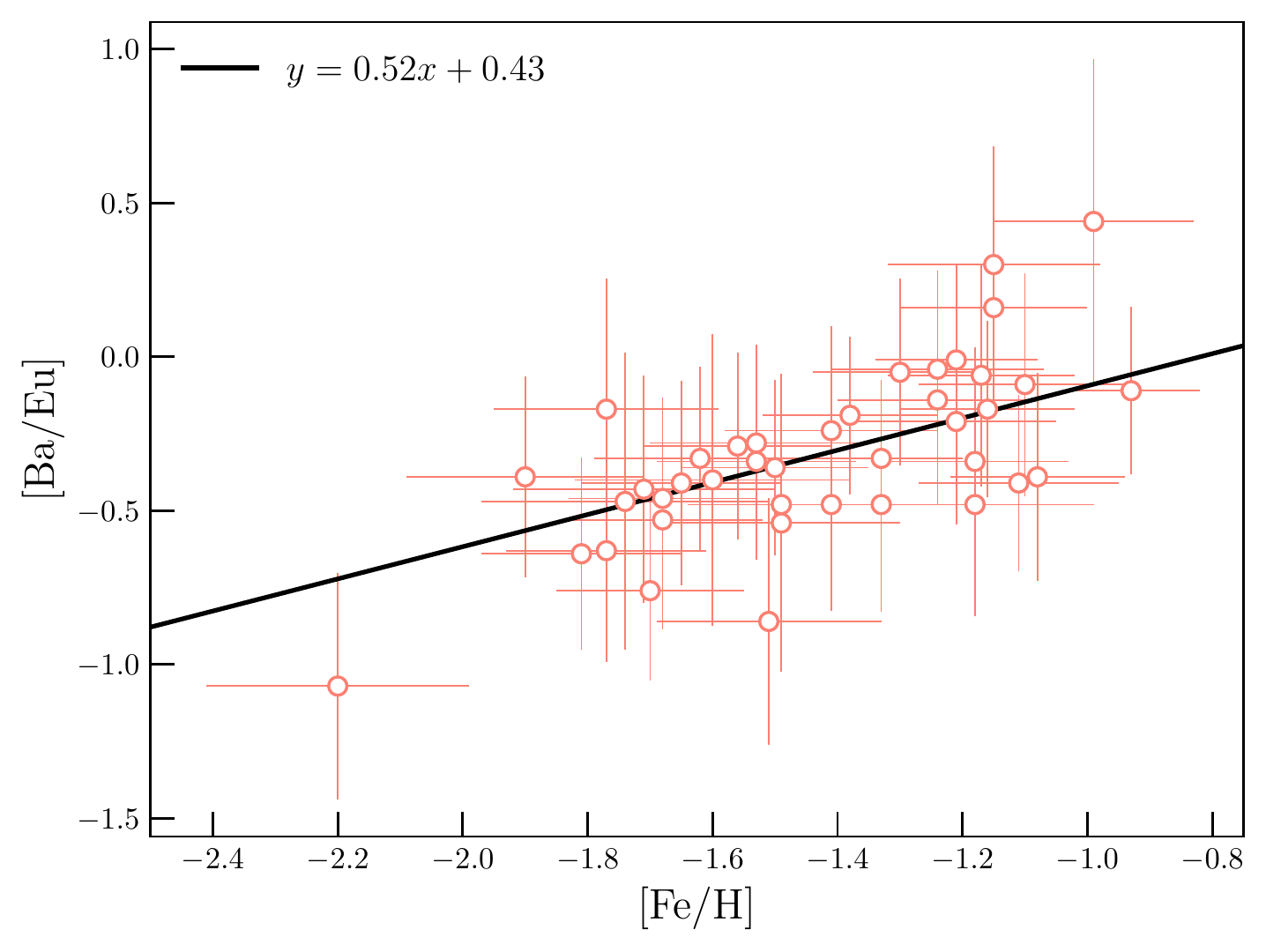}
    \caption{Observed [Ba/Eu] as a function of [Fe/H] from the DART dataset (orange empty points) and the line of best fit, used as a statistical correction to remove $r$-process contributions from [Ba/Fe].}
    \label{fig:baeu}
\end{figure}

Since the number of available Eu measurements is relatively small, we compute a simple statistical correction by fitting a line to [Ba/Eu] as a function of metallicity.
This is shown in Figure~\ref{fig:baeu}, where we derive the best-fit line\footnote{Since measurement uncertainties exist in both the $x$- and $y$-directions, we do this fit by computing $10^{5}$ bootstrap samples. In each sample, we randomly perturb each data point in both the $x$- and $y$-directions, assuming that the true values are distributed normally with standard deviations equal to the measurement errors.
We perform  unweighted linear regression on all samples, and we report the 50th percentile coefficients as the best fit.} $\mathrm{[Ba/Eu]}=0.52\mathrm{[Fe/H]}+0.42$, which can be used to determine the [Ba/Eu] ratio for all stars with metallicity measurements.
The $s$-process-only barium yield, [Ba/Fe]$_{s}$, can then be computed using the fraction $f_{r}$  (Equation~\ref{eq:bacorr}):
\begin{equation}
    \mathrm{[Ba/Fe]}_{s} = \mathrm{[Ba/Fe]} + \log(1-f_{r}).
\end{equation}

Table~\ref{tab:data} lists the full catalog of all abundances, including the original [Ba/Fe] abundances as well as $s$-process-only [Ba/Fe]$_{\mathrm{s}}$. 
The estimated uncertainties include both statistical and systematic uncertainties, as reported in the original catalog papers.
We note that we use [C/Fe] abundances that \citet{Kirby2015} have corrected for astration \citep[the depletion of carbon by stars moving up the red giant branch;][]{Suntzeff1981,Carbon1982,Smith2006} using the corrections proposed by \citet{Placco2014}.

\begin{deluxetable*}{hlhhllllllllll}[t!]
\tablecolumns{14} 
\tablecaption{ Abundance catalog of Sculptor dSph stars. \label{tab:data}} 
\tablehead{ 
\nocolhead{Object} & \colhead{ID} & \nocolhead{RA} & \nocolhead{Dec} & \colhead{[Fe/H]} & \colhead{[Mg/Fe]} & \colhead{[Si/Fe]} & \colhead{[Ca/Fe]} & \colhead{[C/Fe]} & \colhead{[Mn/Fe]} & \colhead{[Ni/Fe]} & \colhead{[Ba/Fe]} & \colhead{$\mathrm{[Ba/Fe]}_{s}^{a}$} &
\colhead{[Eu/Fe]}\\
\nocolhead{} & \colhead{} & \nocolhead{(J2000)} & \nocolhead{(J2000)} & \colhead{(dex)} & \colhead{(dex)} & \colhead{(dex)} & \colhead{(dex)} & \colhead{(dex)} & \colhead{(dex)} & \colhead{(dex)} & \colhead{(dex)} & \colhead{(dex)} & \colhead{(dex)}
}
\rotate
\startdata
\multicolumn{14}{c}{DEIMOS$^{b}$}\\[0.5em]
\tableline
Scl & 1002473 & 00 59 21.58 & -33 42 57.9 & $-2.3\pm0.18$ & $0.65\pm0.93$ & $0.59\pm0.20$ & $0.64\pm0.28$ & \ldots & \ldots & \ldots & \ldots & \ldots & \ldots \\
Scl & 1002447 & 00 59 21.66 & -33 41 1.6 & $-2.04\pm0.16$ & $0.36\pm0.29$ & $0.47\pm0.16$ & $0.33\pm0.19$ & $-0.23\pm0.34$ & \ldots & $-0.19\pm0.29$ & $-0.42\pm0.38$ & $-0.67\pm0.38$ & \ldots \\
Scl & 1002888 & 00 59 23.85 & -33 42 59.1 & $-1.97\pm0.16$ & \ldots & $0.3\pm0.18$ & $0.47\pm0.24$ & \ldots & \ldots & $-0.18\pm0.33$ & \ldots & \ldots & \ldots \\
Scl & 1003386 & 00 59 26.87 & -33 40 28.9 & $-1.3\pm0.15$ & $-0.5\pm0.45$ & $0.21\pm0.25$ & $0.14\pm0.19$ & $-0.25\pm0.33$ & \ldots & $-0.27\pm0.29$ & $-0.34\pm0.39$ & $-0.50\pm0.39$ & \ldots \\
Scl & 1003505 & 00 59 27.13 & -33 43 42.1 & $-1.82\pm0.15$ & $0.26\pm0.20$ & $0.2\pm0.17$ & $0.2\pm0.18$ & $-0.37\pm0.33$ & \ldots & $-0.21\pm0.29$ & $-0.22\pm0.35$ & $-0.47\pm0.35$ & \ldots \\
\tableline
\multicolumn{14}{c}{DART$^{c}$}\\[0.5em]
\tableline
Scl & ET0009 & 01 00 54.18 & -33 40 14.60 & $-1.68\pm0.16$ & $0.57\pm0.22$ & \ldots & $0.2\pm0.07$ & \ldots & $-0.32\pm0.12$ & $-0.02\pm0.15$ & $-0.22\pm0.22$ & $-0.62\pm0.22$ & $0.31\pm0.28$\\
Scl & ET0013 & 01 00 50.24 & -33 36 38.20 & $-1.68\pm0.21$ & $0.53\pm0.27$ & \ldots & $0.28\pm0.10$ & \ldots & $-0.13\pm0.17$ & \ldots & $-0.29\pm0.32$ & $-0.69\pm0.32$ & \ldots\\
Scl & ET0024 & 01 00 34.04 & -33 39 04.61 & $-1.24\pm0.10$ & \ldots & \ldots & $0.0\pm0.14$ & \ldots & $-0.28\pm0.07$ & $-0.39\pm0.13$ & \ldots & \ldots & $-0.17\pm0.22$\\
Scl & ET0026 & 01 00 12.76 & -33 41 16.00 & $-1.8\pm0.16$ & $0.44\pm0.19$ & \ldots & $0.07\pm0.06$ & \ldots & \ldots & $-0.07\pm0.19$ & $-0.24\pm0.21$ & $-0.75\pm0.21$ & \ldots\\
Scl & ET0027 & 01 00 15.37 & -33 39 06.20 & $-1.5\pm0.13$ & $0.12\pm0.15$ & \ldots & $-0.06\pm0.04$ & \ldots & $-0.41\pm0.07$ & $-0.18\pm0.10$ & $-0.19\pm0.19$ & $-0.48\pm0.19$ & \ldots\\
\enddata
\tablenotetext{a}{The $s$-process contribution to [Ba/Fe] is estimated using Equation~\ref{eq:bacorr} as described in the text.}
\tablenotetext{b}{The DEIMOS abundances are compiled from a number of sources: the [Fe/H], [Mg/Fe], [Si/Fe], and [Ca/Fe] measurements from \citet{Kirby2010}; the [C/Fe] measurements from \citet{Kirby2015}; the [Ni/Fe] measurements from \citet{Kirby2018}; the [Mn/Fe] measurements from \citet{delosReyes2020}; and the [Ba/Fe] measurements from \citet{Duggan2018}.}
\tablenotetext{c}{All DART abundances are from \citet{Hill2019}, except the [Mn/Fe] abundances, which are from \citet{North2012}.}
\tablecomments{The errors reported here are total errors (statistical and systematic errors added in quadrature). Only a portion of Table~\ref{tab:data} is shown here; it is published in its entirety (including coordinates) in the machine-readable format online.}
\end{deluxetable*}

\subsection{A Fast, Simple Galactic Chemical Evolution Model} 
\label{sec:gce}

We now consider the simple GCE model used to fit the data described above.
Conceptually this model is similar to that used by \citet{Kirby2011}, and we refer the reader to that work for more details about the individual equations.

The model treats each dwarf galaxy as a chemically homogeneous, open-box system.
In this system, the gas-phase abundance of each element is tracked over a discrete grid of time steps ($\Delta t = 1$~ Myr).
Gas inflows and ejecta from CCSNe, Type Ia SNe, and AGB stars can contribute to the gas-phase abundance of each element, while gas outflows and star formation remove gas-phase elements.
The full model is therefore described by
\begin{eqnarray}
    \xi_{j}(t) & = & \int_{0}^{t}\left(-\dot{\xi}_{j,\mathrm{SF}} + \dot{\xi}_{j,\mathrm{II}} + \dot{\xi}_{j,\mathrm{Ia}} + \dot{\xi}_{j,\mathrm{AGB}} \right. \nonumber\\
    & & + \left. \dot{\xi}_{j,\mathrm{in}} - \dot{\xi}_{j,\mathrm{out}}\right)
\label{eq:model}
\end{eqnarray}
where $\xi_{j}(t)$ is the gas-phase abundance of element $j$.

The other terms in Equation~\ref{eq:model} describe the processes that contribute or remove gas-phase elements.
Star formation is described by a Schmidt-like power law \citep[][]{Schmidt1959,Kennicutt1998}:
\begin{equation}
    \dot{\xi}_{j,\mathrm{SF}}(t) = A_{\star}\left(\frac{M_{j,\mathrm{gas}}(t)}{10^{6}~M_{\odot}}\right)^{\alpha}
    \label{eq:sflaw}
\end{equation}
Following \citet{Kirby2011}, the rate of pristine gas inflows is parameterized with a fast increase and slow decline \citep[e.g.,][]{Lynden-Bell1975}:
\begin{equation}
    \dot{\xi}_{j,\mathrm{in}}(t) = A_{\mathrm{in}}\frac{M_{j,\mathrm{gas}}(0)}{M_{\mathrm{gas}}(0)}\left(\frac{t}{\mathrm{Gyr}}\right)\exp\left(\frac{-t}{\tau_{\mathrm{in}}}\right)
    \label{eq:inflow}
\end{equation}
where $\frac{M_{j,\mathrm{gas}}(0)}{M_{\mathrm{gas}}(0)}$ is the initial mass fraction of element $j$, indicating that the inflows are primordial.\footnote{We assume that the initial mass fractions of H and He are 0.7514 and 0.2486, respectively, from Big Bang nucleosynthesis. All other primordial mass fractions are set to zero.}
We assume that gas outflows are predominantly caused by supernovae, so the outflow is assumed to be linearly proportional to the supernova rate:
\begin{equation}
    \dot{\xi}_{j,\mathrm{out}}(t) = A_{\mathrm{out}}\frac{M_{j,\mathrm{gas}}(t)}{M_{\mathrm{gas}}(t)}\left(\dot{N}_{\mathrm{II}} + \dot{N}_{\mathrm{Ia}}\right)
    \label{eq:outflow}
\end{equation}
From Equations~\ref{eq:sflaw}, \ref{eq:inflow}, and \ref{eq:outflow}, we define the variables $\{A_{\star},\alpha,A_{\mathrm{in}},\tau_{\mathrm{in}},A_{\mathrm{out}}\}$ as free parameters in the model.

\begin{figure}[t!]
    \centering
    \epsscale{1.15}
    \plotone{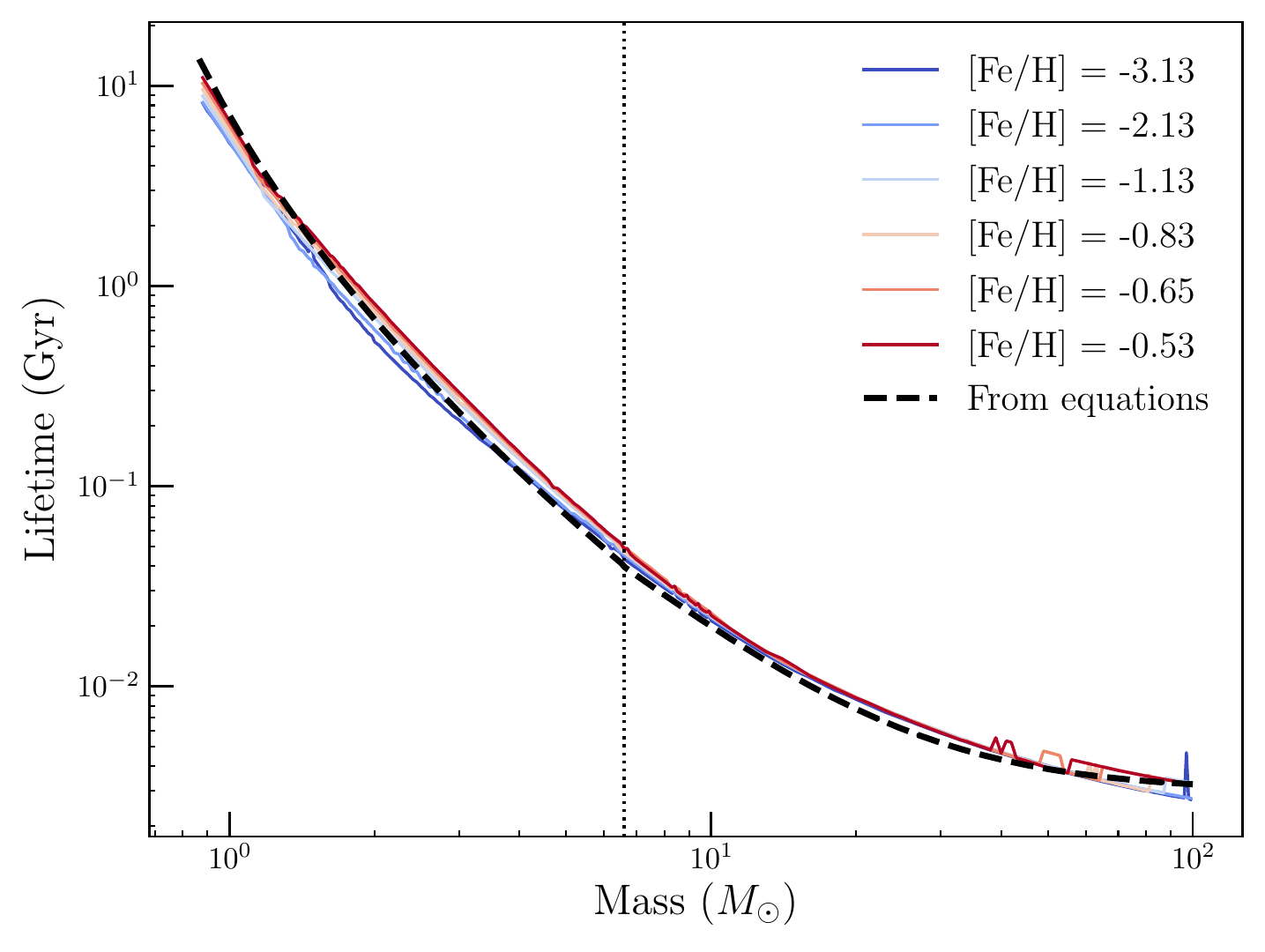}
    \caption{Stellar lifetimes as a function of stellar mass from Equations~\ref{eq:lifetime_himass} and \ref{eq:lifetime_lomass} (black dashed line). The vertical dotted line indicates the domain border of the two equations at $M=6.6~M_{\odot}$. Numerical results from the BPASS stellar evolution code are shown for comparison; bluer (redder) colors represent lower (higher) stellar metallicity.}
    \label{fig:stellarlifetimes}
\end{figure}

The ejecta from SNe and AGB stars at a given time step ($\dot{\xi}_{j,\mathrm{II}}$, $\dot{\xi}_{j,\mathrm{Ia}}$, $\dot{\xi}_{j,\mathrm{AGB}}$) depend on the numbers of SNe/AGB stars occurring at that time, which are determined by convolving the past SFH with a DTD.
For a given type of astrophysical event, the DTD describes the expected event rate as a function of $\tau$, where $\tau$ is the delay time after a $\delta$-function burst of star formation.
The DTD for Type Ia SNe is observed to be a power law with index $\sim-1$ \citep[e.g.,][]{Maoz2010}:
\begin{equation}
    \Psi_{\mathrm{Ia}} = \left\{ \begin{array}{ll}
            0 & t_{\mathrm{delay}} < 0.1~\mathrm{Gyr} \\
            (10^{-3}~\mathrm{Gyr}^{-1}~M_{\odot}^{-1})\left(\frac{t_{\mathrm{delay}}}{\mathrm{Gyr}}\right)^{-1.1} & t_{\mathrm{delay}} \geq 0.1~\mathrm{Gyr}
        \end{array} \right.
    \label{eq:ia_dtd}
\end{equation}
The exact parameterization of the Type Ia DTD is still an open question, and we discuss the effects of modifying the values in Equation~\ref{eq:ia_dtd} on our results later in Section~\ref{sec:inputs}.

For CCSNe and AGB stars, the DTD is primarily set by the stellar IMF, since stellar lifetimes depend strongly on stellar mass.
For ease of comparison with \citet{Kirby2011}, we use a \citet{Kroupa1993} IMF; we consider the effect of changing the IMF in Section~\ref{sec:assumptions}.
We further assume that all stars with birth masses between $10$ and $100~M_{\odot}$ explode as CCSNe at the end of their lifetimes, and that all stars with birth masses $\leq 10~M_{\odot}$ eject mass through AGB winds within the final 1~Myr time step of their lifetimes \citep[][]{Marigo2007}.
Stellar lifetimes are then parameterized as a function of mass using the following equations \citep[][]{Padovani1993}:
\begin{equation}
    \tau_{\star}(M) = (1.2(M/M_{\odot})^{-1.85} + 0.003)~\mathrm{Gyr}
    \label{eq:lifetime_himass}
\end{equation}
for stars with $M>6.6~M_{\odot}$, and
\begin{equation}
    \tau_{\star}(M) = 10^{\frac{0.334-\sqrt{1.790-0.2232(7.764-\log_{10}(M/M_{\odot}))}}{0.1116}}~\mathrm{Gyr}
    \label{eq:lifetime_lomass}
\end{equation}
for stars with $M\leq6.6~M_{\odot}$.
Figure~\ref{fig:stellarlifetimes} compares these equations (black dashed line) with numerical results from the Binary Population and Spectral Synthesis (BPASS) code \citep[][]{Eldridge2017,Stanway2018}, showing that Equations~\ref{eq:lifetime_himass} and \ref{eq:lifetime_lomass} are consistent with stellar evolution models.

To determine the number of SNe/AGB stars at each time step, \citet{Kirby2011} computed the full convolution of the \emph{past} SFH with the DTD (see their Equations 7, 10, and 13). 
In this work, we instead track the elemental abundances in a \emph{forward-looking} array; at each time step, we compute the number of stars that will produce SNe or AGB winds in the future, and we add the nucleosynthetic yields from these SNe/AGB stars (see next section) to the appropriate future times in the array.
This approach, similar to that of the One-zone Model for the Evolution of GAlaxies code \citep[OMEGA;][]{Cote2017}, eliminates most numerical integration from the model. 
Consequently, the computation of the GCE code is approximately an order of magnitude faster, making it possible to run it many times for a Markov Chain Monte Carlo sampler (see Section~\ref{sec:mcmc}).

\begin{deluxetable*}{lll}[t!]
\tablecolumns{5} 
\tablecaption{ Supernova and AGB models. \label{tab:yields}} 
\tablehead{ 
\colhead{Reference} & \colhead{Description}
}
\startdata
\multicolumn{2}{c}{Core-collapse supernova models}\\[0.5em]
\tableline
\citet{Nomoto2013} & Thermal bomb explosions with fixed SN energy ($10^{51}$~erg); \\
& include mass loss, but not rotation; all progenitors explode \\ 
\citet{Limongi2018} & Thermal bomb explosions with variable SN energy; \\
& include mass loss, rotation; only progenitors $\leq25~M_{\odot}$ explode \\
\tableline
\multicolumn{2}{c}{AGB models}\\[0.5em]
\tableline
FRUITY$^{a}$ & Produce $^{13}\mathrm{C}$ pocket with time-dependent convective overshoot; \\
& \citet{Reimers1975} pre-AGB mass loss, \citet{Straniero2006} AGB mass loss\\
Stromlo$^{b}$ & Parameterize $^{13}\mathrm{C}$ pocket with proton abundance profile; \\
& no pre-AGB mass loss, \citet{Vassiliadis1993} AGB mass loss	\\
\tableline
\multicolumn{2}{c}{Type Ia supernova models}\\[0.5em]
\tableline
\citet{Leung2018} & Near-${M}_{\mathrm{Ch}}$ deflagration-to-detonation transition, 2D	\\ 
\citet{Leung2018} & Near-${M}_{\mathrm{Ch}}$ pure deflagration, 2D	\\
\citet{Leung2020} & Sub-${M}_{\mathrm{Ch}}$ ($1.1~M_{\odot}$) double detonation with He shell, 2D \\ 
\citet{Shen2018} & Sub-${M}_{\mathrm{Ch}}$ ($1.1~M_{\odot}$) detonation of bare CO WD, 1D \\	
\enddata
\tablenotetext{a}{This set of yields, described in \citet{Cristallo2011} and \citet{Cristallo2015}, is available at \url{http://fruity.oa-teramo.inaf.it/modelli.pl}.}
\tablenotetext{b}{This set of yields is described in \citet{Lugaro2012}, \citet{Karakas2016}, and \citet{Karakas2018}.}
\end{deluxetable*}

\subsection{Input Nucleosynthetic Yields}
\label{sec:yields}
A key component of our GCE model is the set of nucleosynthetic yields from supernovae and AGB stars. 
A number of models have predicted yield sets, and we summarize a subset of these models in Table~\ref{tab:yields}.
The physical assumptions and computational limitations inherent in these models can produce significant uncertainties in their predicted yields (see Figures~\ref{fig:iasnyields}-\ref{fig:agbyields} in Appendix~\ref{sec:appendix_yields}).

Rather than selecting uncertain yield sets, we instead choose to parameterize the yields with representative analytic expressions, which we will allow to vary in our final fit.
First, we fit analytic functions of mass and metallicity to the yields plotted in Figures~\ref{fig:iasnyields}, \ref{fig:ccsnyields}, and \ref{fig:agbyields}.
For each nucleosynthetic source (CCSNe, AGB stars, and Type Ia SNe), we then vary the input yield sets in the GCE model to determine which elemental abundances are the most sensitive to variations in the yields.
For example, for CCSNe, we run the GCE model twice using the same set of fiducial parameters $\{A_{\mathrm{in}},\tau_{\mathrm{in}},A_{\mathrm{out}},A_{\star},\alpha\}$ \citep[for simplicity, we use the parameters measured by][]{Kirby2011}, varying only the input CCSN yields---either those from \citet{Nomoto2013} or those from \citet{Limongi2018}.
We find that the abundances of C, Mg, and Ca predicted by the GCE model are sensitive to the input CCSN yields---that is, the predicted [X/Fe] abundances change by $>0.2$~dex at the peak of the metallicity distribution function (MDF) ($-1.5<\mathrm{[Fe/H]}<-1.0$).
We therefore define parameters in our analytic functions for C, Mg, and Ca that can be varied to match all of the input yield sets.

Appendix~\ref{sec:appendix_yields} describes the final analytic functions and parameters in more detail.
We obtain seven parameters that represent variations in the nucleosynthetic yields: the Fe yield from Type Ia SNe ($\mathrm{Fe}_{\mathrm{Ia}}$), the exponent of the C yields from CCSNe ($\mathrm{expC}_{\mathrm{II}}$), the normalization of Mg and Ca yields from CCSNe ($\mathrm{normMg}_{\mathrm{II}}$, $\mathrm{normCa}_{\mathrm{II}}$), the normalization of the C yield from AGB stars ($\mathrm{normC}_{\mathrm{AGB}}$), and the normalization and peak of the Ba yield from AGB stars ($\mathrm{normBa}_{\mathrm{AGB}}$, $\mathrm{meanBa}_{\mathrm{AGB}}$).
These parameters are then allowed to vary in the GCE model. 

\section{Results: Dwarf Galaxy Star Formation Histories}
\label{sec:results}

\subsection{Fitting the GCE Model}
\label{sec:mcmc}

We use the chemical evolution model described in the previous section to simultaneously match the MDF as well as the abundance trends of [Mg/Fe], [Si/Fe], [Ca/Fe], [C/Fe], and $s$-process-only [Ba/Fe]$_{\mathrm{s}}$ (see Section~\ref{sec:observations}) as a function of [Fe/H]. 
We choose not to include [Mn/Fe] and [Ni/Fe] in the model fitting.
Manganese and nickel are iron-peak elements that are produced in the same nucleosynthetic events as iron.
However, unlike iron, these elements---particularly manganese---are likely more sensitive to the physics of Type Ia supernova than to the SFH \citep[e.g.,][]{Seitenzahl2013,Seitenzahl2017}.
Rather than use [Mn/Fe] and [Ni/Fe] to fit the model, we instead use these abundances to validate our model and probe additional physics in Section~\ref{sec:mnni}.

\begin{deluxetable*}{lllll}[t!]
\tablecolumns{5} 
\tablecaption{MCMC free parameters. \label{tab:params}} 
\tablehead{ 
\colhead{Parameter} & \colhead{Description} & \colhead{Prior} & \colhead{Initial value} & \colhead{Best-fit result}
}
\startdata
$A_{\mathrm{in}}$ & Normalization of gas infall rate ($10^{9}~M_{\odot}~\mathrm{Gyr}^{-1}$) & $\mathcal{U}(0,5)$ & 1.07 & $0.53^{+0.09}_{-0.08}$\\
$\tau_{\mathrm{in}}$ & Gas infall time constant (Gyr) & $\mathcal{U}(0,1)$ & 0.16 & $0.27^{+0.03}_{-0.02}$\\
$A_{\mathrm{out}}$ & Gas lost per supernova ($M_{\odot}~\mathrm{SN}^{-1}$) & $\mathcal{U}(0,20)$ & 4.01 & $4.79^{+0.19}_{-0.19}$\\
$A_{\star}$ & Normalization of star formation law ($10^{6}~M_{\odot}~\mathrm{Gyr}^{-1}$) & $\mathcal{U}(0,10)$ & 0.89 & $0.79^{+0.14}_{-0.13}$\\
$\alpha$ & Power-law index of star formation law & $\mathcal{U}(0,2)$ & 0.82 & $0.72^{+0.09}_{-0.07}$\\
Fe$_{\mathrm{Ia}}$ & Fe yield from Type Ia SNe & $\mathcal{U}(0,0.9)$ & 0.80 & $0.58^{+0.03}_{-0.03}$\\
expC$_{\mathrm{II}}$ & Exponent of C yield from CCSNe & $\mathcal{U}(0,2)$ & 1.0 & $1.32^{+0.03}_{-0.03}$\\
normMg$_{\mathrm{II}}$ & Normalization of Mg yield from CCSNe & $\mathcal{U}(0,2)$ & 1.0 & $1.41^{+0.10}_{-0.09}$\\
normCa$_{\mathrm{II}}$ & Normalization of Ca yield from CCSNe & $\mathcal{U}(0,0.5)$ & 0.01 & $0.24^{+0.05}_{-0.05}$\\
normC$_{\mathrm{AGB}}$ & Normalization of C yield from AGB stars & $\mathcal{U}(0.4,5)$ & 0.60 & $1.98^{+0.44}_{-0.36}$\\
normBa$_{\mathrm{AGB}}$ & Normalization of Ba yield from AGB stars & $\mathcal{U}(0,1)$ & 0.33 & $1.08^{+0.31}_{-0.21}$\\
meanBa$_{\mathrm{AGB}}$ & Mass of peak Ba yield from AGB stars ($M_{\odot}$) & $\mathcal{N}(2,0.25)$ & 2.0 & $2.80^{+0.16}_{-0.16}$
\enddata
\tablecomments{For all parameters, the best-fit values are reported as the median (50th percentile) values, with uncertainties based on the 16th and 84th percentiles.}
\end{deluxetable*}

Following \citet{Kirby2011}, we treat the chemical
evolution model as tracing a path $\epsilon_{j}(t)$ in the six-dimensional \{[Fe/H], [Mg/Fe], [Si/Fe], [Ca/Fe], [C/Fe], [Ba/Fe]$_{\mathrm{s}}$\}-space.
The probability of a star forming at any time $t$ is given by $dP/dt=\dot{M_{\star}}(t)/M_{\star}$, where $M_{\star}$ is the final stellar mass.
The likelihood of a star $i$ forming along the path defined by the chemical evolution model is therefore given by the line integral of $dP/dt$ along the path $\epsilon_{j}$ for the total duration of the model $t_{\mathrm{final}}$:
\begin{equation}
    L_{i} = \int_{0}^{t_{\mathrm{final}}}\left(\prod_{j}\frac{1}{\sqrt{2\pi}\sigma_{i,j}}\exp\frac{-(\epsilon_{i,j}-\epsilon_{j}(t))^2}{2(\sigma_{i,j})^2}\right)\frac{\dot{M_{\star}}(t)}{M_{\star}}dt
\end{equation}
where $\epsilon_{i,j}$ is the $j$th observed elemental abundance ratio for star $i$, and $\sigma_{i,j}$ is the corresponding uncertainty.
The final time step, $t_{\mathrm{final}}$ is not a free parameter.
Rather, it is the last time step before the galaxy runs out of gas.

The total likelihood for a model is therefore proportional to the product of $L_{i}$ for all $N$ stars:
\begin{eqnarray}
    L & = & \prod_{i}^{N}L_{i} \times \left(\frac{1}{\sqrt{2\pi}\delta M_{\star,\mathrm{obs}}}\exp\frac{-(M_{\star,\mathrm{obs}}-M_{\star,\mathrm{model}})^2}{2(\delta M_{\star,\mathrm{obs}})^2}\right. \nonumber\\ 
    & & \left.\times \frac{1}{\sqrt{2\pi}\delta M_{\mathrm{gas},\mathrm{model}}}\exp\frac{-(M_{\mathrm{gas},\mathrm{model}})}{2(\delta M_{\mathrm{gas},\mathrm{obs}})}
    \right)^{0.1N}
    \label{eq:likelihood}
\end{eqnarray}
Here, the additional terms require that the final stellar and gas masses of the model ($M_{\star,\mathrm{model}}$ and $M_{\mathrm{gas},\mathrm{model}}$) match the observed masses ($M_{\star,\mathrm{obs}}$ and $M_{\mathrm{gas},\mathrm{obs}}$) within the observational uncertainties.
The exponent $0.1N$ is chosen to weight these terms relative to the abundance distributions, to prevent them from dominating the likelihood while ensuring that the models end up with approximately the correct stellar and gas masses.
We use observed stellar masses and uncertainties from \citet{Woo2008}.
We assume the observed gas mass $M_{\mathrm{gas},\mathrm{obs}}$ is zero for all dSphs, and we choose an arbitrary uncertainty of $\delta M_{\mathrm{gas},\mathrm{obs}} = 10^3~M_{\odot}$ to ensure the model converges. 
This is similar to observed upper limits on gas measurements \citep[e.g.,][who find an upper limit of $(3.2\pm0.4)\times10^{3}~M_{\odot}$ on Sculptor's HI mass]{Putman2021}.

To estimate the values of the 12 free parameters $\{A_{\mathrm{in}},\tau_{\mathrm{in}},A_{\mathrm{out}},A_{\star},\alpha,\mathrm{Fe}_{\mathrm{Ia}},\mathrm{expC}_{\mathrm{II}},\mathrm{normMg}_{\mathrm{II}},\\\mathrm{normCa}_{\mathrm{II}},\mathrm{normC}_{\mathrm{AGB}},\mathrm{normBa}_{\mathrm{AGB}},\mathrm{meanBa}_{\mathrm{AGB}}\}$ that minimize the negative log-likelihood ($-\ln L$), we used the \texttt{emcee} Python module \citep{Foreman-Mackey2013} to implement a Markov Chain Monte Carlo (MCMC) ensemble sampler.
Table~\ref{tab:params} describes the inputs and outputs of this MCMC sampling: the priors, initial values from linear optimization, and the best-fit values for each parameter.

For all free parameters except $\mathrm{meanBa}_{\mathrm{AGB}}$, we assumed uniform priors  with lower limits at 0 to avoid unphysical negative values and upper limits chosen based on the range of parameters determined by \citet{Kirby2011}.
For the parameter $\mathrm{meanBa}_{\mathrm{AGB}}$, which dictates the mass of the AGB stars that produce the most barium, we assume a normal prior with a mean of $2~M_{\odot}$ and standard deviation $0.5~M_{\odot}$.
This is because $\mathrm{meanBa}_{\mathrm{AGB}}$ describes when the $s$-process begins to contribute meaningfully to the abundance of barium: low $\mathrm{meanBa}_{\mathrm{AGB}}$ means that lower-mass AGB stars produce most of the $s$-process barium, so [Ba/Fe]$_{s}$ begins to increase at later times and higher [Fe/H].
As a result, any measurements of [Ba/Fe]$_{s}$ at low [Fe/H] will have outsized leverage on the value of $\mathrm{meanBa}_{\mathrm{AGB}}$.
We therefore enforce a Gaussian prior to keep this parameter within physically reasonable limits.
Initial values of the parameters were chosen by performing a simple linear optimization of $-\ln L$.
We sampled $10^6$ steps using 32 ensemble members or ``walkers'' initialized about these values, discarded the first $10^{4}$ ``burn-in'' steps, and used the remaining steps to sample the posterior distribution of the parameters.

\subsection{The Star Formation History of Sculptor dSph}

\begin{figure}[t!]
    \centering
    \epsscale{1.22}
    \plotone{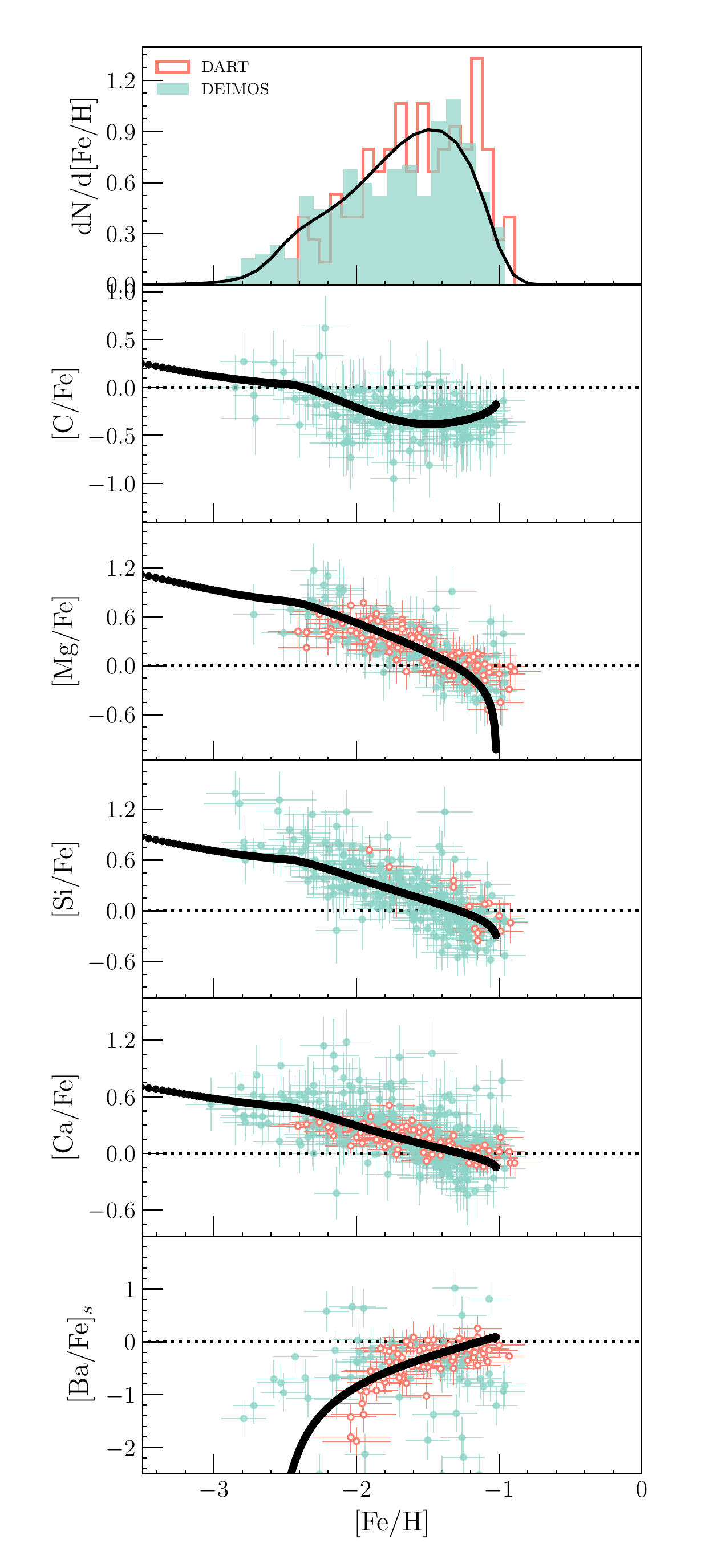}
    \caption{Metallicity distribution function (top panel) and abundance trends as a function of [Fe/H] from the best-fit GCE model for Sculptor dSph (black lines). Filled blue (empty orange) histogram and points represent the observed data from DEIMOS (DART). Note that [Fe/H] from DART and [Ba/Fe] from DEIMOS were excluded from fitting and are shown here for illustration.}
    \label{fig:gceresult_scl}
\end{figure}

We use the GCE model described in the previous section to fit the stellar abundances of Sculptor dSph.
For Mg, Si, Ca, and C, we simultaneously fit our model to both the medium-resolution abundances from DEIMOS (Table~\ref{tab:yields}) and the high-resolution DART abundances \citep[][]{Hill2019,North2012} (filled blue and empty orange points, respectively, in Figure~\ref{fig:gceresult_scl}).
When fitting [Fe/H], we use only the DEIMOS yields because the DART sample is significantly smaller than the DEIMOS sample ($N_{\mathrm{DART}}=89$ compared to $N_{\mathrm{DEIMOS}}=376$).
When fitting $s$-process Ba abundances, we use only the DART yields because the statistical correction used to remove the $r$-process contribution (Section~\ref{sec:ba}) was based on the DART yields of Eu and Ba.

\begin{figure}[t!]
    \centering
    \epsscale{1.15}
    \plotone{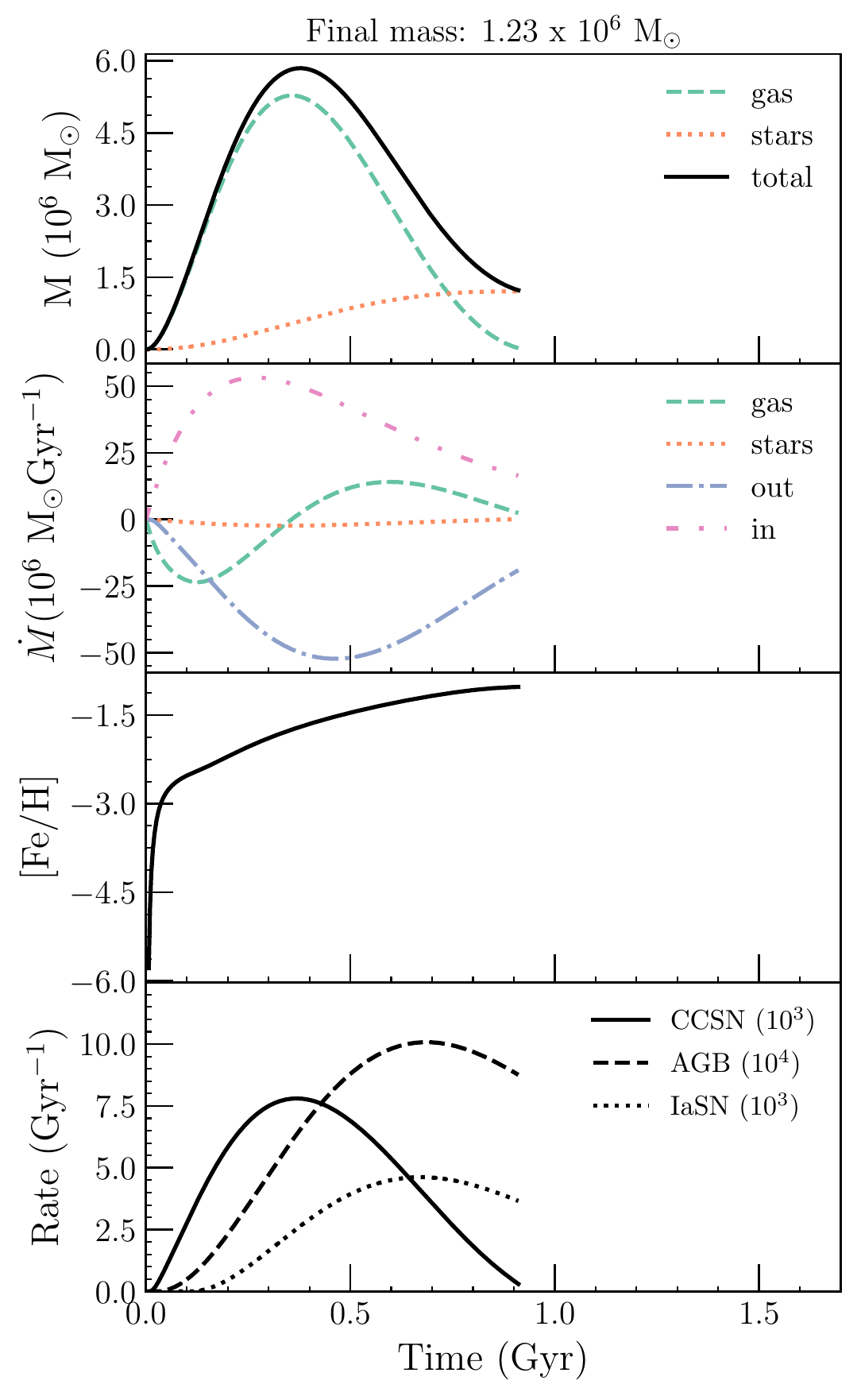}
    \caption{Outputs from the best-fit GCE model as a function of time. Top: the total masses of the different components. Second from top: the changes in mass. Third from top: the overall stellar metallicity. Bottom: the rates of nucleosynthetic events.}
    \label{fig:gceoutput_scl}
\end{figure}

Figure~\ref{fig:gceresult_scl} shows the best-fit abundance trends from the GCE model, illustrating that the model fits the stellar abundances reasonably well.
We also plot the outputs from the best-fit GCE model, showing how the components of the galaxy change over time, in Figure~\ref{fig:gceoutput_scl}.
We note that while the rates of Type Ia SNe (dotted line in bottom panel) appear relatively low---particularly when compared to AGB stars at late times---this is because the average yield per Type Ia SN is much larger (typically by at least two orders of magnitude; see Figures~\ref{fig:iasnyields} and \ref{fig:agbyields}) than the average yield per AGB wind. 
Despite their small numbers, Type Ia SNe dominate galactic chemical evolution at late times.

\begin{figure*}[t!]
    \centering
    \epsscale{0.9}
    \plotone{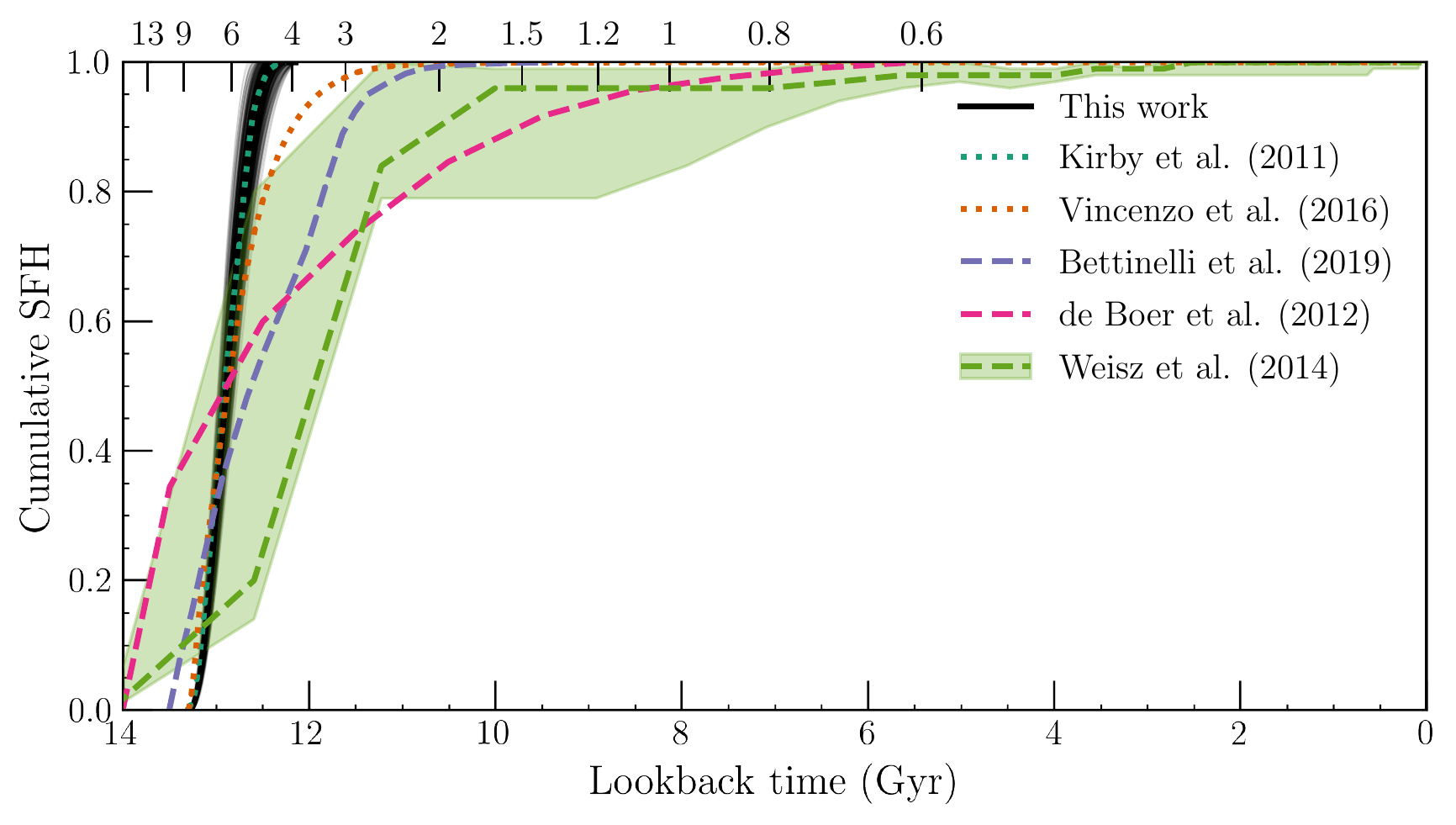}
    \caption{Cumulative star formation history from the best-fit GCE model for Sculptor dSph (black solid lines). Photometrically derived SFHs from the literature are plotted as dashed lines and shaded regions, while SFHs derived from GCEs and stellar abundances are plotted as dotted lines.}
    \label{fig:sfh_scl}
\end{figure*}

The corresponding cumulative SFH from this best-fit model is shown in Figure~\ref{fig:sfh_scl} as a thick black line; to give a rough sense of the uncertainty in this model, thin black lines represent random realizations of the posterior distribution.
We find that Sculptor has an ancient stellar population: the best-fit SFH is a single burst of star formation with a relatively short duration of $\sim0.92$~Gyr.
We note that chemical evolution models measure relative rather than absolute ages, so the exact location of our measured SFH on the time axis is uncertain.
\citet{Bromm2011} point out that the halos thought to be the hosts of the first galaxies are predicted to form $\sim500$~Myr after the Big Bang; we therefore assume our SFH begins when the universe is $500$~Myr old.

\section{Discussion}
\label{sec:discussion}

\subsection{Comparison to Previous Literature}
\label{sec:literature}
As shown in Figure~\ref{fig:sfh_scl}, we also compare our results to previous measurements of dSph SFHs.
Nearly all measurements of Sculptor's SFH have been based on CMD fitting.
\citet{DaCosta1984} and \citet{Monkiewicz1999} fit a few model isochrones of varying ages and metallicities to CMDs and found that most of the stars in Sculptor must be relatively ancient ($\sim13\pm2$~Gyr old).
\citet{Dolphin2002} reanalyzed the data from \citet{Monkiewicz1999} by interpolating over a grid of synthetic CMDs to get a ``true model CMD'' rather than fitting discrete isochrones, coming to the same conclusion that Sculptor must be ``entirely ancient.''
On the other hand, \citet{deBoer2012} argued that Sculptor has a much more extended SFH than these previous measurements would indicate. 
Using a new technique to simultaneously fit the CMD and the MDF, \citet{deBoer2012} found that Sculptor had a continuous period of star formation with a duration of $\sim6-7$~Gyr (dashed pink line in Figure~\ref{fig:sfh_scl}).
\citet{Savino2018} extended this technique to include horizontal branch stars in fitting the CMD and MDF, finding an extended SFH with a prominent tail of star formation at younger ages.

However, these results were based on relatively shallow photometry that did not reach significantly below the main-sequence turnoff (MSTO).
Other measurements from deeper photometry appear to confirm the original picture of Sculptor as an ancient galaxy that formed all of its stars in a short burst.
\citet{Weisz2014} measured CMDs using the Hubble Space Telescope, obtaining photometry with $\sim30\%$ completeness at $\sim2$~mag below the MSTO\@.
From these, they determined that Sculptor formed $\sim90\%$ of its stars $>10$~Gyr ago (dashed green line and shaded region in Figure~\ref{fig:sfh_scl}).
Similarly, \citet{Bettinelli2019} used DECam to obtain a CMD down to $\sim2$~mag below the MSTO and found that Sculptor had a single burst of star formation with a full width at half maximum of $\sim2.2$~Gyr (dashed purple line in Figure~\ref{fig:sfh_scl}).

As Figure~\ref{fig:sfh_scl} shows, SFHs measured from GCE models are also ancient, in qualitative agreement with the results from deep photometry.
Yet GCE models tend to produce shorter absolute star formation durations (dotted lines) than the photometrically derived SFHs (dashed lines).
This discrepancy likely arises from the limitations of photometric methods; although photometry is excellent at determining absolute ages, as discussed in Section~\ref{sec:intro} the age resolution of CMD fitting degrades for old and metal-poor populations.
Abundance-derived estimates of the star formation duration may therefore better resolve the \emph{relative} spread in ages within the ancient population of Sculptor.

Our results are largely consistent with previous SFH measurements using GCE models.
The SFH we derive in this work (solid black line in Figure~\ref{fig:sfh_scl}) has a duration of $0.92$~Gyr.
This is slightly shorter than the SFH found by \citet{Vincenzo2016}, who used a one-zone model and found that 99\% of the stars in Sculptor dSph formed within the first $2.16$~Gyr of its evolution (dotted orange line in Figure~\ref{fig:sfh_scl}).
However, their model only aimed to fit the stellar metallicity distribution function and did not include information about individual abundances.
\citet{Kirby2011}, on the other hand, used similar methods to our work and traced both the stellar MDF and several abundance ratios.
They found that Sculptor finished forming stars within $1.1$~Gyr; as Figure~\ref{fig:sfh_scl} shows, this SFH (dotted blue line) is entirely consistent with the random realizations of the posterior distributions from our model (black lines).

This work expands on that of \citet{Kirby2011} by using more data---they used only DEIMOS abundances and did not use any [C/Fe] or [Ba/Fe] abundance information.
Our model incorporates $s$-process abundances, as well as additional abundances measured from the VLT DART survey.
It also has additional free parameters beyond those used in \citet{Kirby2011}, which we use to fit the analytic functions describing nucleosynthetic yields.
The consistency between our results and those of \citet{Kirby2011} suggests that including these additional parameters in our model does not significantly impact our main results; instead, as we will discuss later in Section~\ref{sec:ccsnagb}, these parameters provide additional useful information about predicted nucleosynthetic yield sets.

\subsection{Model Assumptions}
\label{sec:assumptions}

\begin{figure}[t!]
    \centering
    \epsscale{1.1}
    \plotone{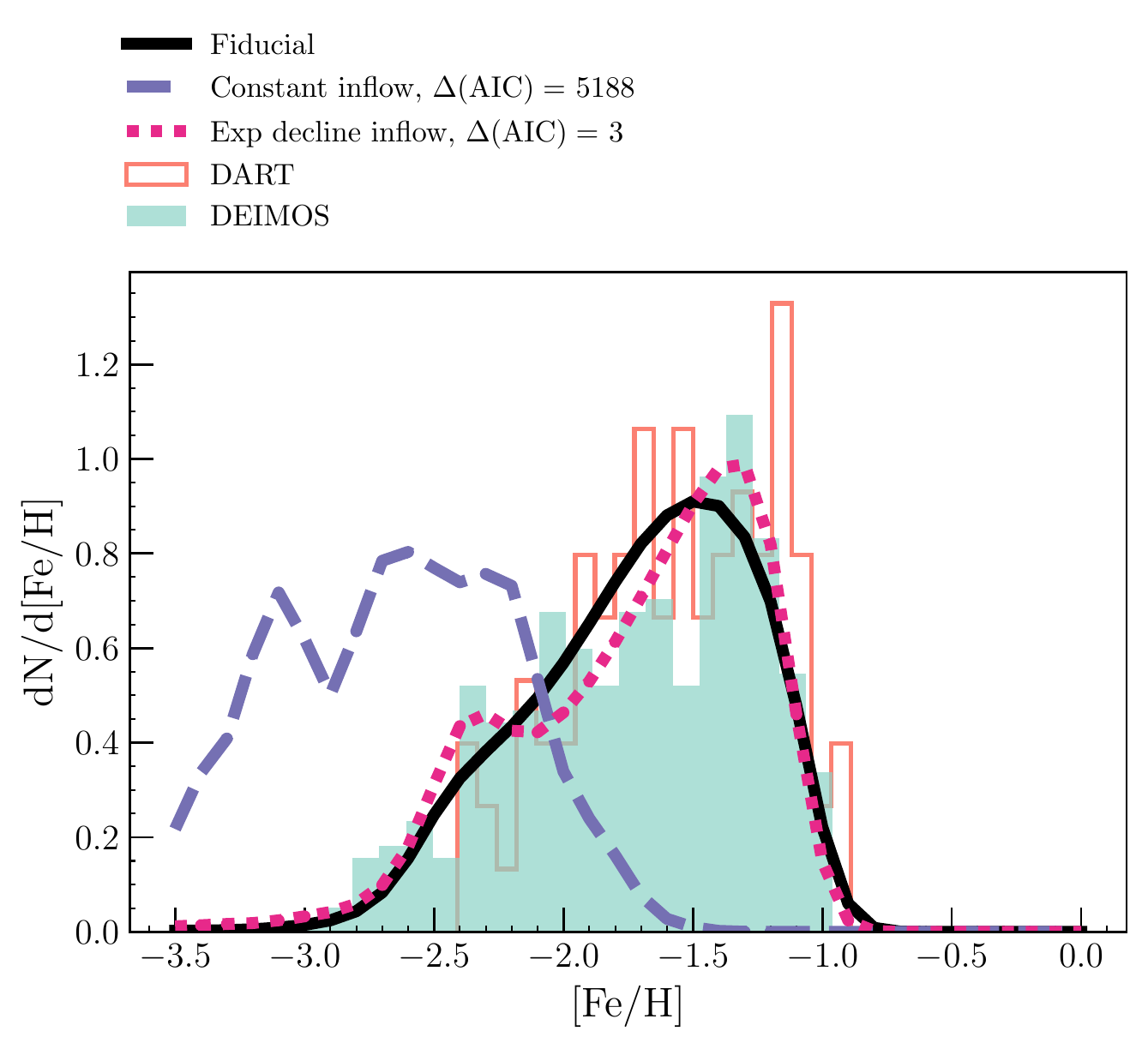}
    \caption{Comparisons between the MDFs of Sculptor from the fiducial GCE model (black solid line) and from models with other gas inflow parameterizations. The assumed parameterization of gas inflow significantly influences the shape of the MDF.}
    \label{fig:inflowtest}
\end{figure}

\begin{figure}[h!]
    \centering
    \epsscale{1.1}
    \plotone{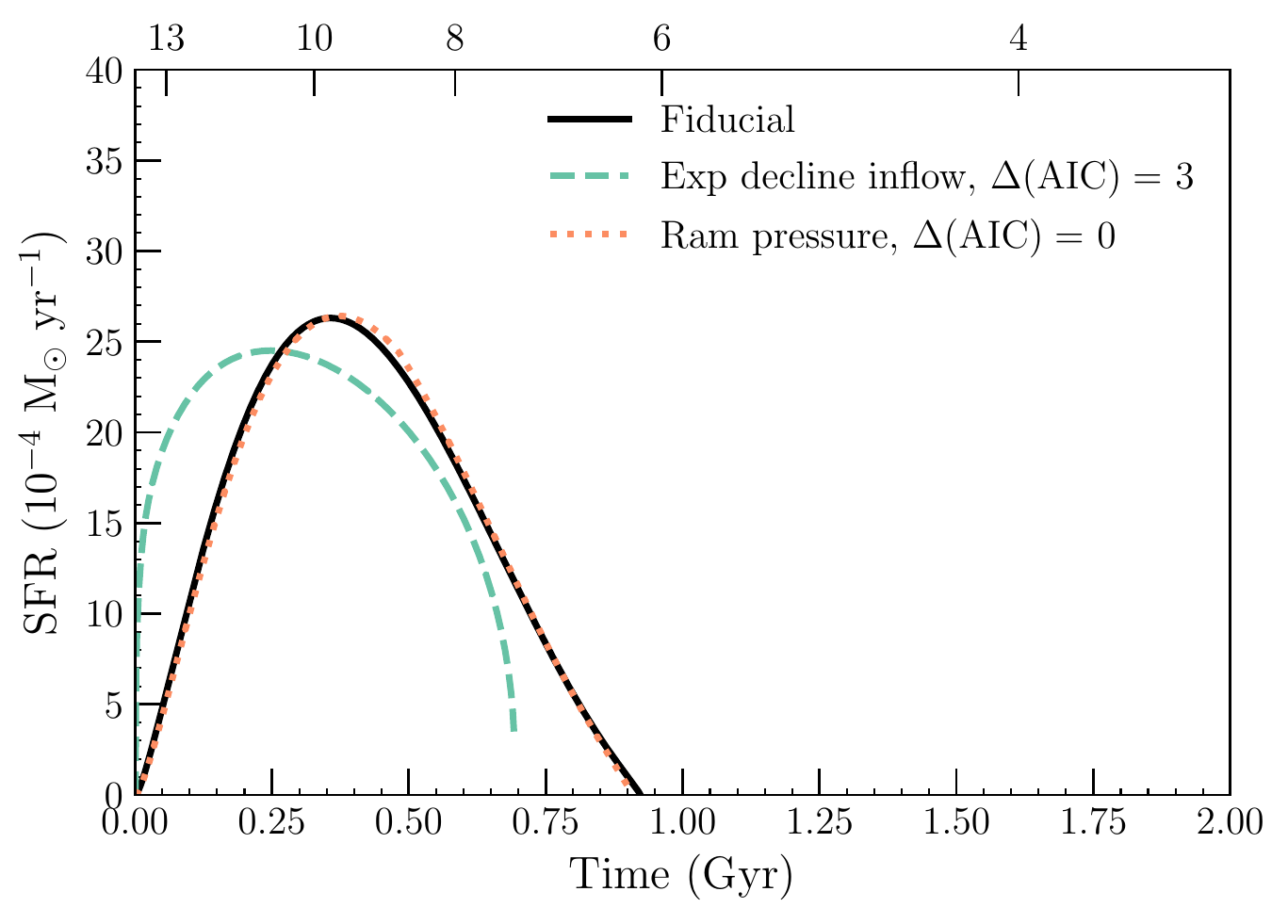}
    \plotone{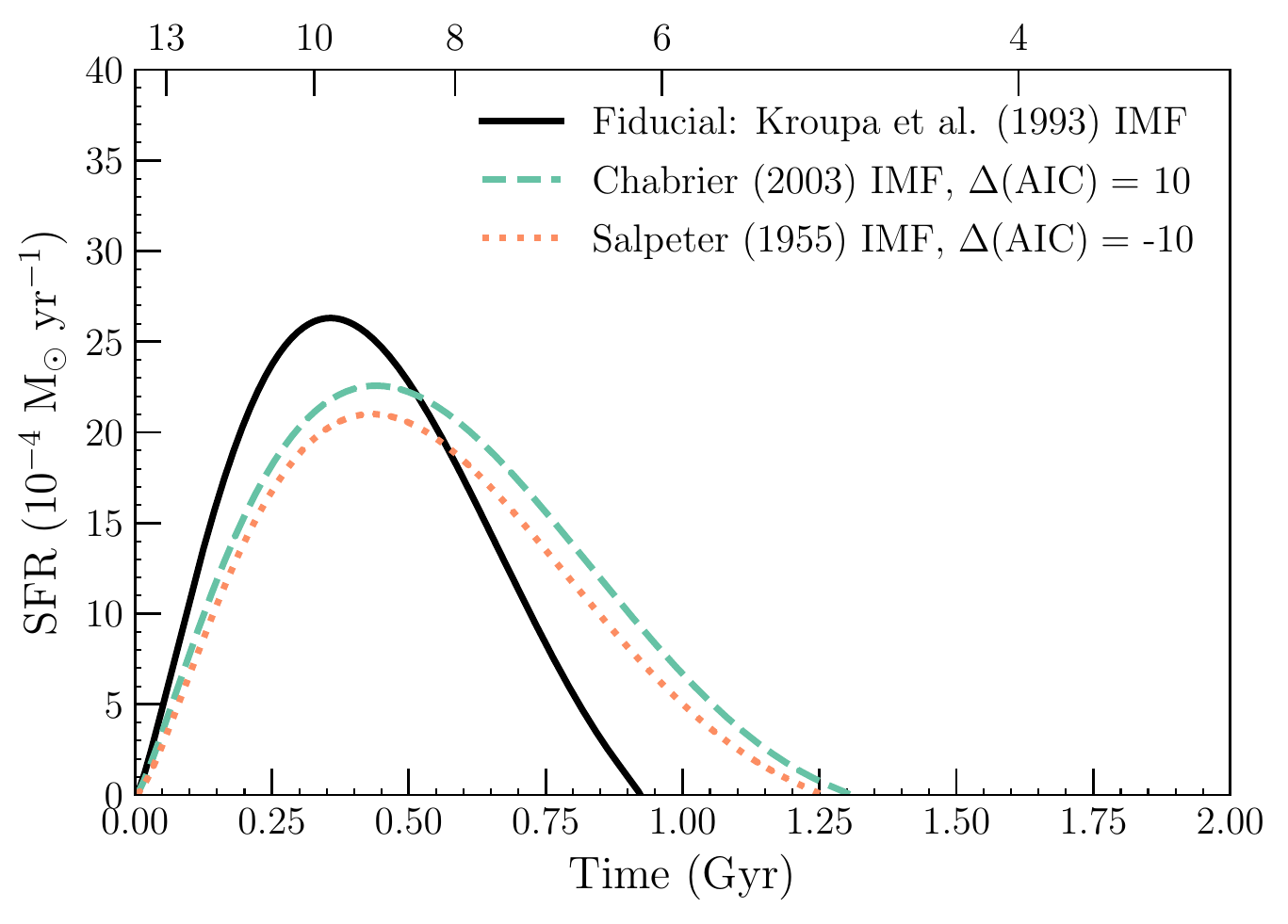}
    \plotone{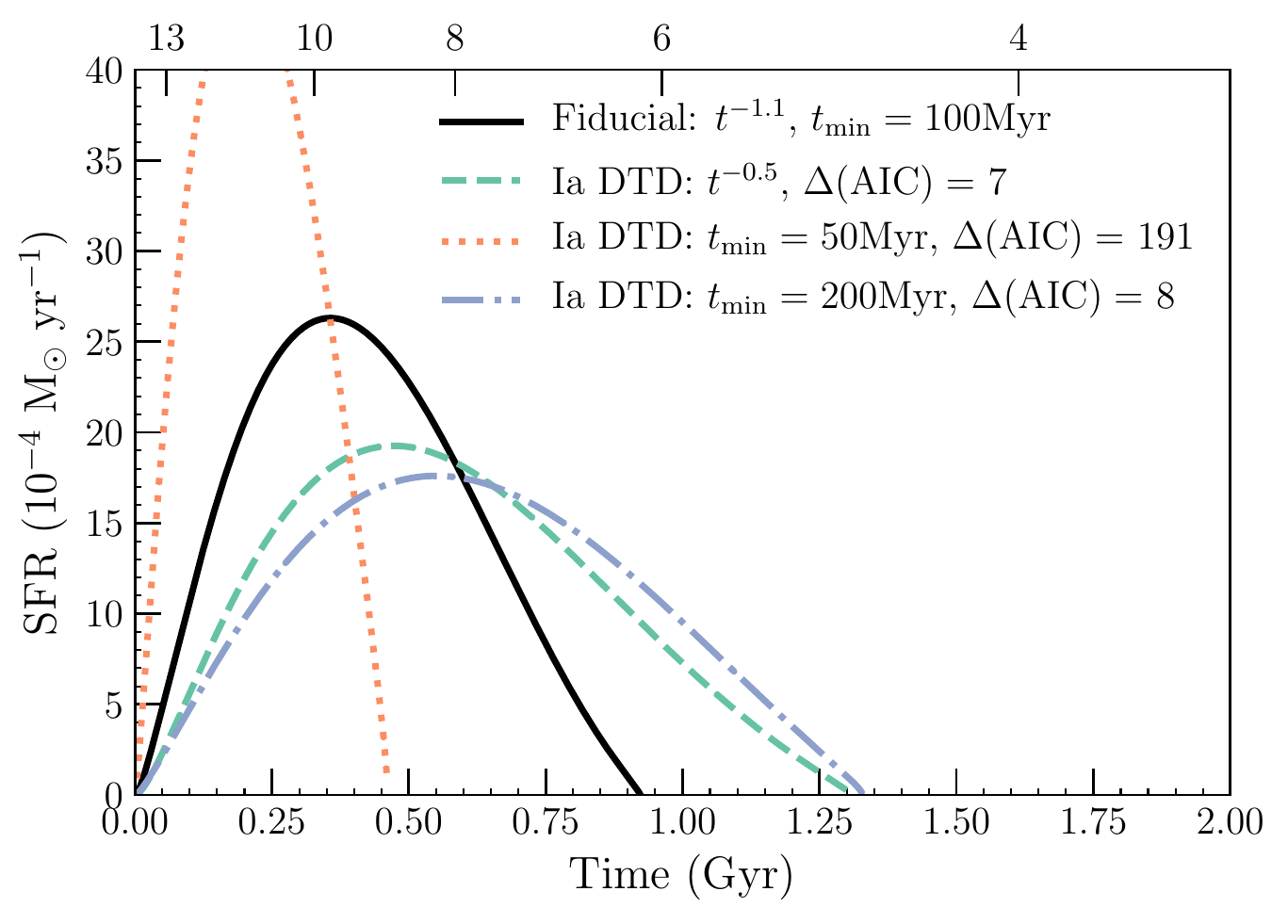}
    \caption{Comparisons between the SFHs of Sculptor from the fiducial GCE model (black solid line) and from other test models as described in the text. Top: results from varying assumptions in model construction: using an exponentially declining gas inflow (green dashed line) and adding ram pressure stripping (orange dotted line). Middle: results from varying the stellar IMF: \citet{Chabrier2003} IMF (green dashed line) and \citet{Salpeter1955} IMF (orange dotted line). Bottom: results from varying Type Ia DTD: decreasing the power law index to $-0.5$ (green dashed line), decreasing the minimum delay time to 50~Myr (orange dotted line), and increasing the minimum delay time to 200~Myr (purple dot-dashed line).}
    \label{fig:sfhtest}
\end{figure}

In this section, we discuss the simplifying assumptions on which our chemical evolution model depends, and their potential impact on our results.
These assumptions can broadly be classified into three categories: assumptions inherent to the construction of the model, assumptions in the model inputs, and other potential sources of systematic errors.
In Figures~\ref{fig:inflowtest} and \ref{fig:sfhtest}, we illustrate some of the effects of changing these assumptions.

\subsubsection{Model construction}
In constructing our simple one-zone model, we have made a number of inherent assumptions.
For example, the primary assumption in our model is the ``one-zone'' assumption of instantaneous mixing. 
This approximation is reasonably well-founded for dSphs; \citet{Escala2018} found that in simulated dwarf galaxies, a well-mixed ISM due to turbulent metal diffusion successfully reproduces observed abundance distributions of dSphs.

Other model assumptions may have greater impacts on the measured SFH.
For example, the parameterization of gas inflow (Equation~\ref{eq:inflow}) strongly influences the shape of the SFH because star formation depends on the gas mass, which is in turn predominantly set by the gas inflow.
Fortunately, the inflow parameterization is constrained by the metallicity distribution function.
This is shown in Figure~\ref{fig:inflowtest}, which illustrates the output MDFs of models using other common inflow parameterizations.
To compare the goodness-of-fit of these models, we compute the Akaike information criterion (AIC); a lower AIC implies less information loss, so a given model is a ``better'' fit than our fiducial model if $\Delta(\mathrm{AIC})=\mathrm{AIC}_{\mathrm{model}}-\mathrm{AIC}_{\mathrm{fiducial}}$ is positive. 
A constant gas inflow, given by the equation
\begin{equation}
    \dot{\xi}_{j,\mathrm{in}}(t) = A_{\mathrm{in}}\frac{M_{j,\mathrm{gas}}(0)}{M_{\mathrm{gas}}(0)}
    \label{eq:inflow_const}
\end{equation} 
where $A_{\mathrm{in}}$ is a free parameter, is entirely unable to match the MDF of Sculptor dSph ($\Delta(\mathrm{AIC})\gg 0$). 

An exponentially declining inflow,
\begin{equation}
    \dot{\xi}_{j,\mathrm{in}}(t) = A_{\mathrm{in}}\frac{M_{j,\mathrm{gas}}(0)}{M_{\mathrm{gas}}(0)}\exp\left(\frac{-t}{\tau_{\mathrm{in}}}\right)
    \label{eq:inflow_expdec}
\end{equation}
with free parameters $A_{\mathrm{in}}$ and $\tau_{\mathrm{in}}$, performs better than the constant inflow model, but not as well as our fiducial model ($\Delta(\mathrm{AIC})=3$).
As shown in the top panel of Figure~\ref{fig:sfhtest}, using the exponentially declining inflow produces a single burst of star formation that ends after 0.69~Gyr (green dashed line).
This is shorter than our fiducial model SFH (total duration of 0.92~Gyr) by a factor of $\sim25\%$, likely because higher gas inflow at early times accelerates star formation and more quickly depletes gas.
While this demonstrates that the parameterization of the gas inflow may have a significant effect on the predicted shape and duration of the SFH, more complex parameterizations of the gas inflow are somewhat disfavored for Sculptor dSph.
As noted by \citet{deBoer2012} and \citet{Escala2018}, among others, a bursty SFH might produce a wider spread in [$\alpha$/Fe] at fixed [Fe/H] than observed in Sculptor \citep[although see, e.g.,][]{Marcolini2008}.

Our model also ignores environmental effects like tidal or ram pressure stripping, which may cut off gas inflows and/or contribute to the removal of gas.
Simulations \citep[e.g.,][]{Kazantzidis2017} have shown that these environmental effects can, depending on galaxy orbital parameters, contribute significantly to gas loss in dSphs.
To test this, we apply a simple model of ram pressure stripping by adding a constant to the gas outflow (Equation~\ref{eq:outflow}). 
This parameterization is similar to that used in the analytic model of \citet{Kirby2013}, who found that ram pressure stripping successfully reproduced the metallicity distribution function of Sculptor dSph.
Following this parameterization, we apply an additional constant outflow starting at $\mathrm{[Fe/H]}=-1.5$ \citep[based on the best-fit parameterization found by][who found that ram pressure stripping in Sculptor began at $\mathrm{[Fe/H]}<-1.41$]{Kirby2013}:
\begin{equation}
    \dot{\xi}_{j,\mathrm{out}}(t) = \frac{M_{j,\mathrm{gas}}(t)}{M_{\mathrm{gas}}(t)}\left[A_{\mathrm{out}}\left(\dot{N}_{\mathrm{II}} + \dot{N}_{\mathrm{Ia}}\right) + C_{\mathrm{ram}}\right].
    \label{eq:outflow_stripping}
\end{equation}
Here, we allow $C_{\mathrm{ram}}$ to be a free parameter setting the removal of gas due to ram pressure stripping.
We fit it along with the other parameters in our GCE model,\footnote{When fitting $C_{\mathrm{ram}}$, we assume a uniform prior of $\mathcal{U}(0,5)$ and an initial value of 0.} finding a best-fit value of $C_{\mathrm{ram}}=2.76^{+1.53}_{-1.77}~M_{\odot}~\mathrm{yr}^{-1}$.
The resulting SFH is shown as a orange dotted line in the top panel of Figure~\ref{fig:sfhtest}, which is almost exactly the same as the fiducial model; the overall star-forming duration of 0.90~Gyr is $\sim2\%$ different from that predicted by the fiducial model and has almost the same goodness-of-fit (as measured by the AIC).

There are a number of other model assumptions that may affect our measurement of the SFH.
For example, we do not account for reionization, which may heat infalling gas and delay star formation. 
Furthermore, our parameterization of the outflow rate as linearly proportional to the supernova rate (Equation~\ref{eq:outflow}) assumes that supernovae are the only factors in determining outflow rates and that all supernovae (both Type Ia and core-collapse) contribute equally to outflows.
However, recent hydrodynamic simulations indicate that other factors, such as a galaxy's gas fraction (e.g., M. Orr et al., in preparation) and the clustering of supernovae \citep[e.g.,][]{Fielding2018}, can strongly affect whether supernovae are able to produce galactic outflows. 
Equation~\ref{eq:outflow} also assumes that the galactic gravitational potential remains constant, but dark matter accretion, environmental effects, and stellar feedback might all affect the underlying gravitational potential of a galaxy.
Fully addressing these assumptions would require a more sophisticated model, which we defer to future work.

\subsubsection{Model inputs}
\label{sec:inputs}
We now consider the effects of different inputs in our GCE model.
First, we consider the assumed forms of the supernova and AGB DTDs.
For CCSNe and AGB stars, these are set by a combination of the stellar IMF and stellar lifetimes.
Our initial model assumes a \citet{Kroupa1993} IMF\@.
The middle panel of Figure~\ref{fig:sfhtest} shows the effect of using a \citet{Chabrier2003} or \citet{Salpeter1955} IMF on the output SFH\@.
Both these IMFs are slightly steeper than our fiducial model, producing more low-mass stars and fewer high-mass stars.
The dearth of massive stars means it takes longer for stellar feedback to remove gas from the galaxy, leading to longer predicted SFHs.
Adopting a \citet{Chabrier2003} IMF leads to a total star formation duration of $1.30$~Gyr (a $41\%$ increase from the fiducial duration of $0.92$~Gyr), while the \citet{Salpeter1955} IMF predicts a star formation duration of $1.25$~Gyr (a $36\%$ increase).

Finally, perhaps the most uncertain input in our model is the Type Ia DTD.
Although it appears that many observational studies have reached a consensus on the power law index of the Type Ia DTD of $\sim-1$ \citep[see, e.g., the review by][]{Maoz2012}, for completeness we consider the effect of changing the DTD index.
We find that a shallower DTD power law ($\propto t^{-0.5}$) provides a fit that is only slightly worse than our fiducial model ($\Delta(\mathrm{AIC})=7$).
This shallower DTD flattens the rate of Type Ia SNe as a function of time, so that it takes longer for Type Ia SNe to remove gas from the galaxy.
As a result, the best-fit SFH from this shallower DTD (green dashed line in the bottom panel of Figure~\ref{fig:sfhtest}) is longer than the fiducial SFH by $42\%$ (an increase from 0.92 to 1.31~Gyr).

We next consider the minimum delay time $t_{\mathrm{min}}$.
This parameter, which sets the time at which Type Ia SNe first ``turn on,'' is currently poorly constrained by observations \citep[e.g.,][]{Maoz2012}.
A number of studies \citep[e.g.,][]{Greggio2005,Castrillo2021,Wiseman2021} suggest $t_{\mathrm{min}}$ should be as short as $\sim0.04$~Gyr, which is approximately the main-sequence lifetime of the most massive secondary in the binary system that produces a Type Ia SN (i.e., an $8~M_{\odot}$ progenitor).
Our fiducial GCE model (Section~\ref{sec:results}) assumes a minimum delay time of $t_{\mathrm{min}}\sim0.1$~Gyr, which corresponds to the formation of a carbon-oxygen white dwarf from a $4-5~M_{\odot}$ progenitor.
We test the effect of making Type Ia SNe more prompt by decreasing $t_{\mathrm{min}}$ to 0.05~Gyr.
As shown by the orange dotted line in the bottom panel of Figure~\ref{fig:sfhtest}, this decreases the duration of the SFH by $50\%$ and provides a significantly worse fit to the data ($\Delta(\mathrm{AIC})=191$).

We also consider a later minimum delay time, which may imply a lower-mass progenitor.
Increasing $t_{\mathrm{min}}$ to 0.2~Gyr flattens the rate of Type Ia SNe; similar to the shallower $t^{-0.5}$ power law, this increases the duration of the SFH by $50\%$ (purple dotted-dashed line in the bottom panel of Figure~\ref{fig:sfhtest}).
The fit to the data is only slightly worse than the fiducial model ($\Delta(\mathrm{AIC})=9$).
\citet{Kirby2011} also experimented with increasing the delay time to 0.3~Gyr.
With this change, they found a much longer SFH duration of 3.7~Gyr for Sculptor.
Our result is different because we have the benefit of AGB products (C and Ba) to better constrain the SFH\@.

We conclude that changing the Type Ia DTD to reduce the number of Type Ia SNe may produce models that can also fit the observed data reasonably well.
These models may produce SFHs that are more extended than our fiducial model by up to $\sim50\%$.
We note, however, that full constraints on the form of the Type Ia DTD are beyond the scope of this paper, and we defer this to a later work (M. de los Reyes et al., in preparation).

\subsubsection{Additional systematics}
Another potential source of systematic error is the selection effect of the observed stellar population.
The DEIMOS spectroscopic sample was centrally concentrated to maximize the number of member stars per slit mask.
However, evidence suggests that there are two distinct stellar populations in Sculptor: a kinematically cold, relatively metal-rich centrally concentrated population, and a warm, metal-poor spatially extended population \citep[e.g.,][]{Battaglia2008}.
Indeed, although they do not agree on absolute ages in Sculptor's SFH, previous studies that measure the SFH in different regions of Sculptor find that the duration of star formation is longer in its more central regions.

Both \citet{deBoer2012} and \citet{Bettinelli2019} define a ``central'' region of $\sim10\arcmin$ in which the SFH duration is longest.
This region has roughly the same radial extent as our spectroscopic sample \citep[both the DEIMOS and DART samples are concentrated in the inner $\sim12\arcmin$ of Sculptor dSph, as shown in Figure 2 of][]{Hill2019}, so we expect our measurements to probe the younger, more metal-rich stars.
In this case, the duration of star formation for the overall stellar population in Sculptor may be even shorter than the $\sim1$~Gyr duration of star formation that we measure.
We also note that the existence of a bimodal population suggests multiple bursts of star formation, but no measurement of Sculptor's SFH---including our own---finds evidence of more than one burst of star formation.

\section{Implications for Nucleosynthetic Yields}
\label{sec:nucleosynthesis}

Not only does our GCE model result in a robust measurement of the SFH, it can also probe the nucleosynthetic production of different elements.
The free parameters used to describe nucleosynthetic yields can be used to compare different sets of theoretical yields.
Furthermore, even though we do not fit [Mn/Fe], [Ni/Fe], total [Ba/Fe], and [Eu/Fe] in our initial GCE model, our best-fit model can still be used to provide insight into their nucleosynthesis.

\subsection{Comparing CCSN and AGB yield sets}
\label{sec:ccsnagb}

As described in Section~\ref{sec:yields}, the nucleosynthetic yields from CCSN and AGB models can vary widely. 
We aimed to work around these uncertainties by parameterizing the yields with analytic functions (Appendix~\ref{sec:appendix_yields}).
For a number of these yields we defined free parameters, allowing the model to vary the yields in order to best match the observed abundance trends.
By examining the behavior of these free parameters, we can determine whether certain yield sets are preferred over others.

Among the CCSN yields, we varied the yields for C, Mg, and Ca. 
The best-fit yields, shown as blue lines in Figure~\ref{fig:ccsnyields}, are a better match to the yields compiled by \citet{Nomoto2013} rather than those from the models of \citet{Limongi2018}.
There are a number of differences between these two yield sets that could contribute to this discrepancy.
At least one major difference is in the assumed explosion ``landscape'' (i.e., what masses of progenitor stars explode): \citet{Limongi2018} assumed that all stars with initial masses above $25~M_{\odot}$ implode, so that no yields are produced from explosive nucleosynthesis.
As a result, many of their predicted CCSN yields (the dotted lines in Figure~\ref{fig:ccsnyields}) are zero for high-mass progenitors.
Our model, which finds nonzero yields at masses $>25~M_{\odot}$ (i.e., extremely prompt nucleosynthesis) for the parameterized elements C, Mg, and Ca, is therefore more consistent with the \citet{Nomoto2013} yields, which assume that all massive progenitor stars explode.
This may suggest that an explosion landscape different from the simple one assumed by \citet{Limongi2018} is needed to match observations \citep[see, e.g.,][]{Griffith2021}.

Our GCE model also exhibits other features---high Mg yields and relatively low Ca yields at low progenitor masses---in the best-fit yields that appear to be consistent with the \citet{Nomoto2013} yields.
These features could result from a number of model assumptions.
For example, \citet{Limongi2018} evolve massive stars from pre-main sequence to pre-supernova, fix the explosion energy such that all exploding models produce exactly $0.1~M_{\odot}$ of Fe, and include the effects of rotation.
\citet{Nomoto2013}, on the other hand, evolve massive stars from pre-supernova to explosion, assume fixed supernova energy $E_{\mathrm{SN}}=10^{51}~\mathrm{erg}$, and do not include rotation.
Determining the exact reasons why our model appears to agree with the C, Mg, and Ca yields of \citet{Nomoto2013} is beyond the scope of this paper.
We also note that our result differs slightly from the result of \citet{Nunez2021}, who independently infer CCSN nucleosynthetic yields for a number of elements using damped Ly$\alpha$ systems.
They find that while the C yields from \citet{Nomoto2013}\footnote{Specifically, Nu\~{n}ez et al. consider the yields from \citet{Kobayashi2006} and \citet{Nomoto2006}, from which \citet{Nomoto2013} built their yield catalog.} are consistent with the observed [C/Fe] ratio, the \citet{Limongi2018} yields are typically more consistent with other observed abundance ratios, including [C/O].

For the AGB yields, we varied the yields of C and Ba, shown in Figure~\ref{fig:agbyields}.
The best-fit yields (blue lines) are more consistent with the Stromlo AGB yields from \citet{Karakas2016} and \citet{Karakas2018}, which predict enhanced C and Ba values at low progenitor masses.
The AGB yields predicted by the FRUITY models \citep[][]{Cristallo2015} do not predict such a large enhancement.
The large C and Ba yields may result from the Stromlo models including a deeper third dredge-up mixing, which brings more He-shell material to the stellar surface.
This mixing predominantly affects elements made by neutron capture such as barium, along with the products of partial He-shell burning such as carbon.

\begin{figure}[t!]
    \centering
    \epsscale{1.15}
    \plotone{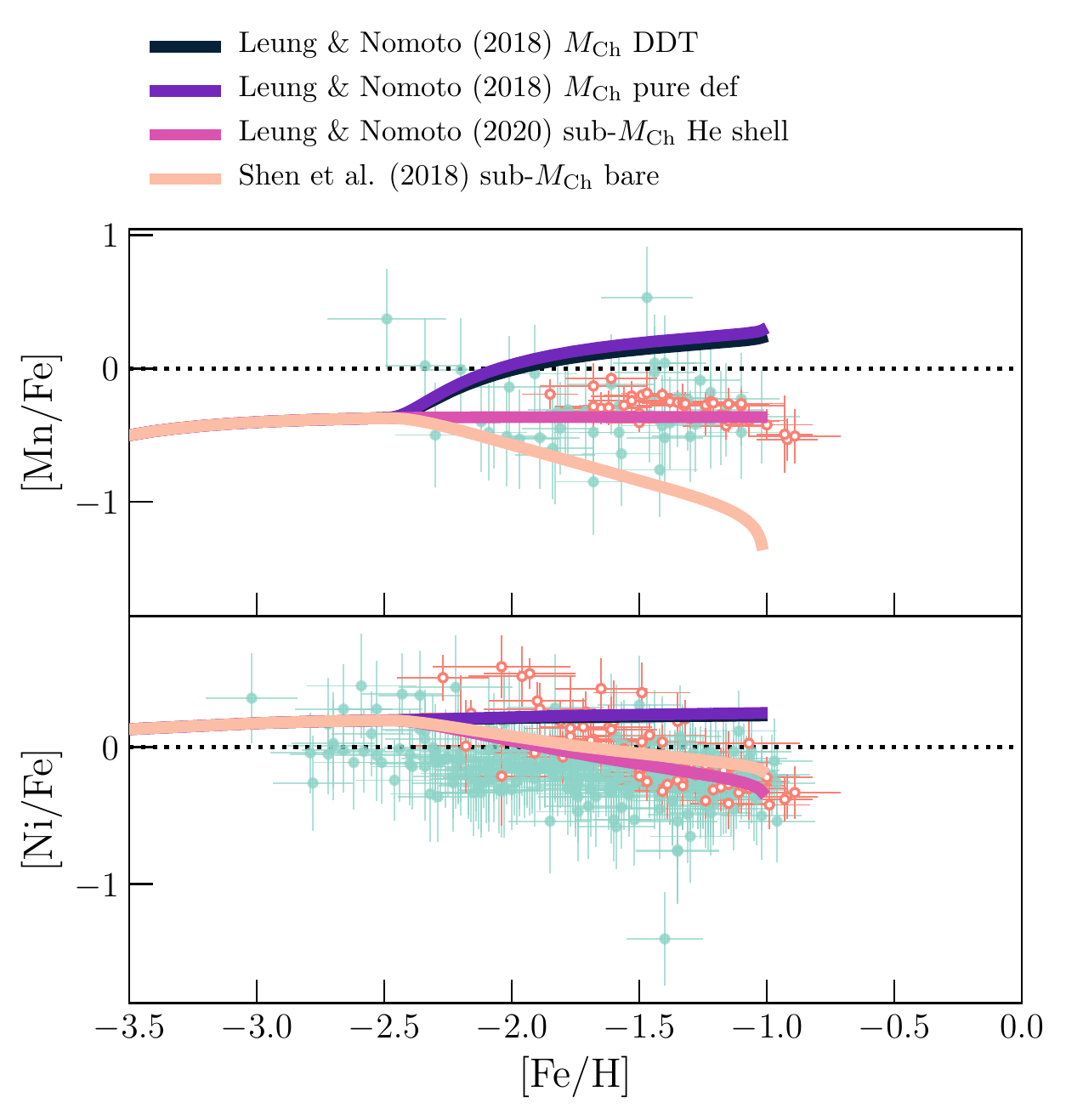}
    \caption{Comparisons among different Type Ia yields in the best-fit GCE model (solid lines) for manganese (top) and nickel (bottom). Points denote observations from DEIMOS (filled blue points) and DART (empty orange points).}
    \label{fig:iacomparison}
\end{figure}

\subsection{Probing Type Ia SNe Using Mn and Ni}
\label{sec:mnni}

Manganese and nickel are both iron-peak elements predominantly produced in Type Ia SNe.
As discussed in Section~\ref{sec:mcmc}, Mn is particularly sensitive to the physics of Type Ia SNe.
The production of the only stable isotope of manganese, $^{55}$Mn, depends strongly on the density---and therefore the mass---of the progenitor white dwarf \citep[][]{Seitenzahl2013,Seitenzahl2017}.
White dwarfs near the Chandrasekhar mass ($M_{\mathrm{Ch}}\approx1.4~M_{\odot}$) are expected to produce solar or supersolar [Mn/Fe] and [Ni/Fe], while sub-$M_{\mathrm{Ch}}$ models tend to produce subsolar [Mn/Fe] and [Ni/Fe] (see, e.g., Figure~\ref{fig:iasnyields}).

\citet{Kirby2019} and \citet{delosReyes2020} found that in Sculptor dSph, subsolar [Ni/Fe] and [Mn/Fe] abundances indicate that sub-$M_{\mathrm{Ch}}$ Type Ia SNe likely dominate.
Both found that the Ni and Mn yields in Sculptor were most consistent with $\sim1~M_{\odot}$ sub-$M_{\mathrm{Ch}}$ models from \citet{Leung2020}.
However, these comparisons were based on an analytic model that made a number of simplifying assumptions (e.g., that CCSNe are the only nucleosynthetic sources at early times, that CCSN yields are metallicity-independent, that the only contributions to stellar abundances are from supernovae).
Our GCE model can be used to make more sophisticated comparisons.

In Figure~\ref{fig:iacomparison}, we use the best-fit parameters from Table~\ref{tab:params} and vary the Type Ia yields of Mn and Ni.
We find that the observed [Mn/Fe] and [Ni/Fe] trends are consistent with sub-$M_{\mathrm{Ch}}$ models, supporting the hypothesis that sub-$M_{\mathrm{Ch}}$ Type Ia SNe likely dominate in Sculptor dSph.
In particular, the \citet{Leung2020} model of a $\sim1~M_{\odot}$ CO white dwarf with a helium shell appears to best fit our data.
This is in broad agreement with the findings of \citet{Kobayashi2020}, who used a one-zone chemical evolution model to study dSphs and found that a significant fraction of sub-$M_{\mathrm{Ch}}$ Type Ia SNe are needed to reproduce the observed iron-peak abundances. 
Similar results have also been obtained in the Milky Way \citep[see, e.g.,][and references therein]{Palla2021}.

Although we have attempted to pick a representative subset of Type Ia models, there are many other yield sets that we do not consider here.
For the models that we have used in our comparison, we have included the effect of metallicity-dependent yields; however, our GCE model does not make any assumptions about the masses of Type Ia SNe that explode, so we are unable to include mass-dependent yields.
For simplicity, we also do not consider the effects of multiple simultaneous channels of Type Ia SNe.  In contrast, see, e.g., \citet{Kobayashi2020} who found that Type Iax SNe, potentially the pure deflagrations of hybrid C+O+Ne white dwarfs, are required to match the observed $\mathrm{[Mn/Fe]}$ abundances.
Furthermore, we do not attempt to correct the observed Mn or Ni abundances for the effects of stellar atmospheres not being in local thermodynamic equilibrium (LTE). These non-LTE effects may be especially significant for manganese \citep[e.g.,][]{Bergemann2019}.

\begin{figure*}[t!]
    \centering
    \epsscale{1.12}
    \plottwo{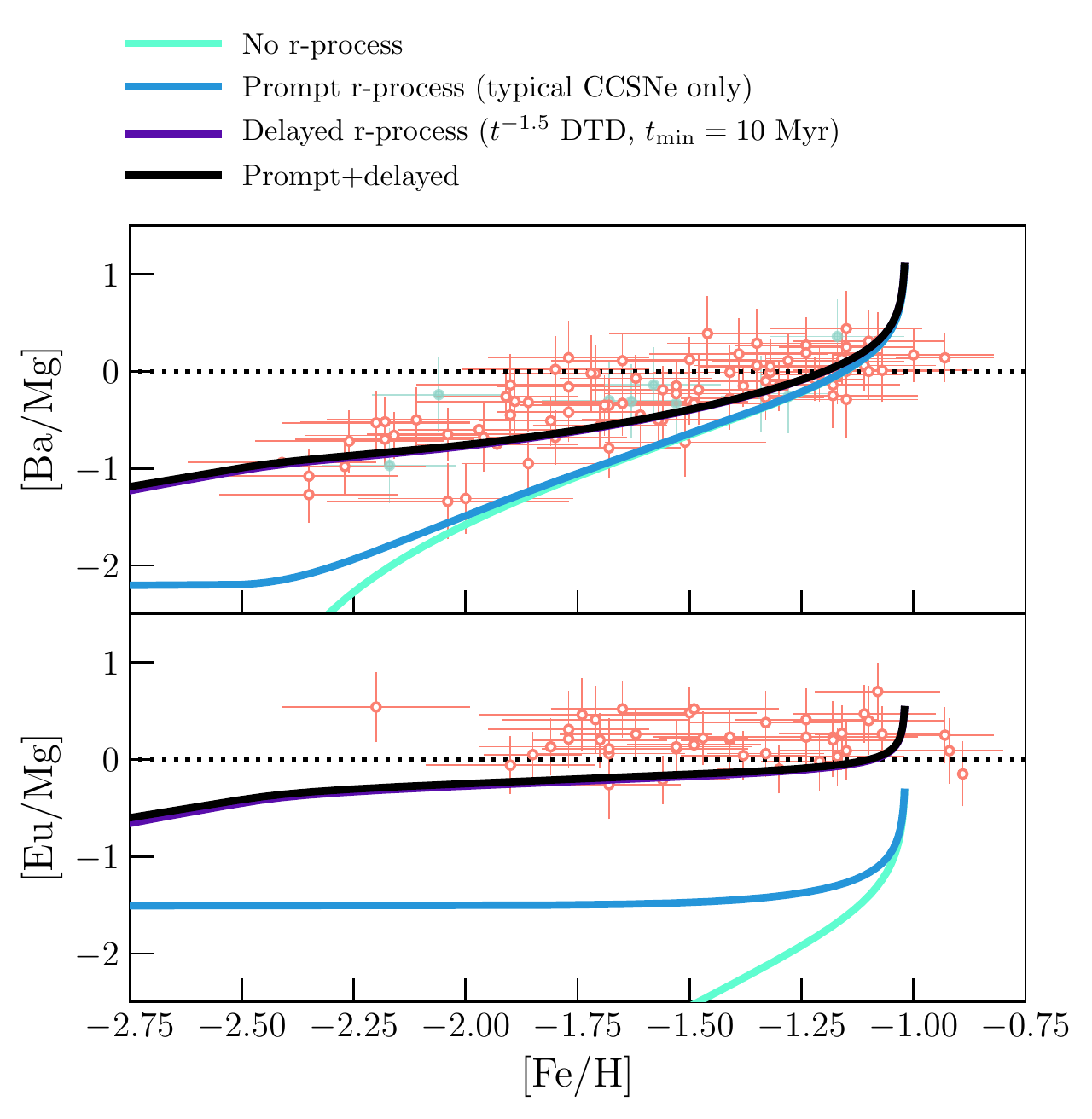}{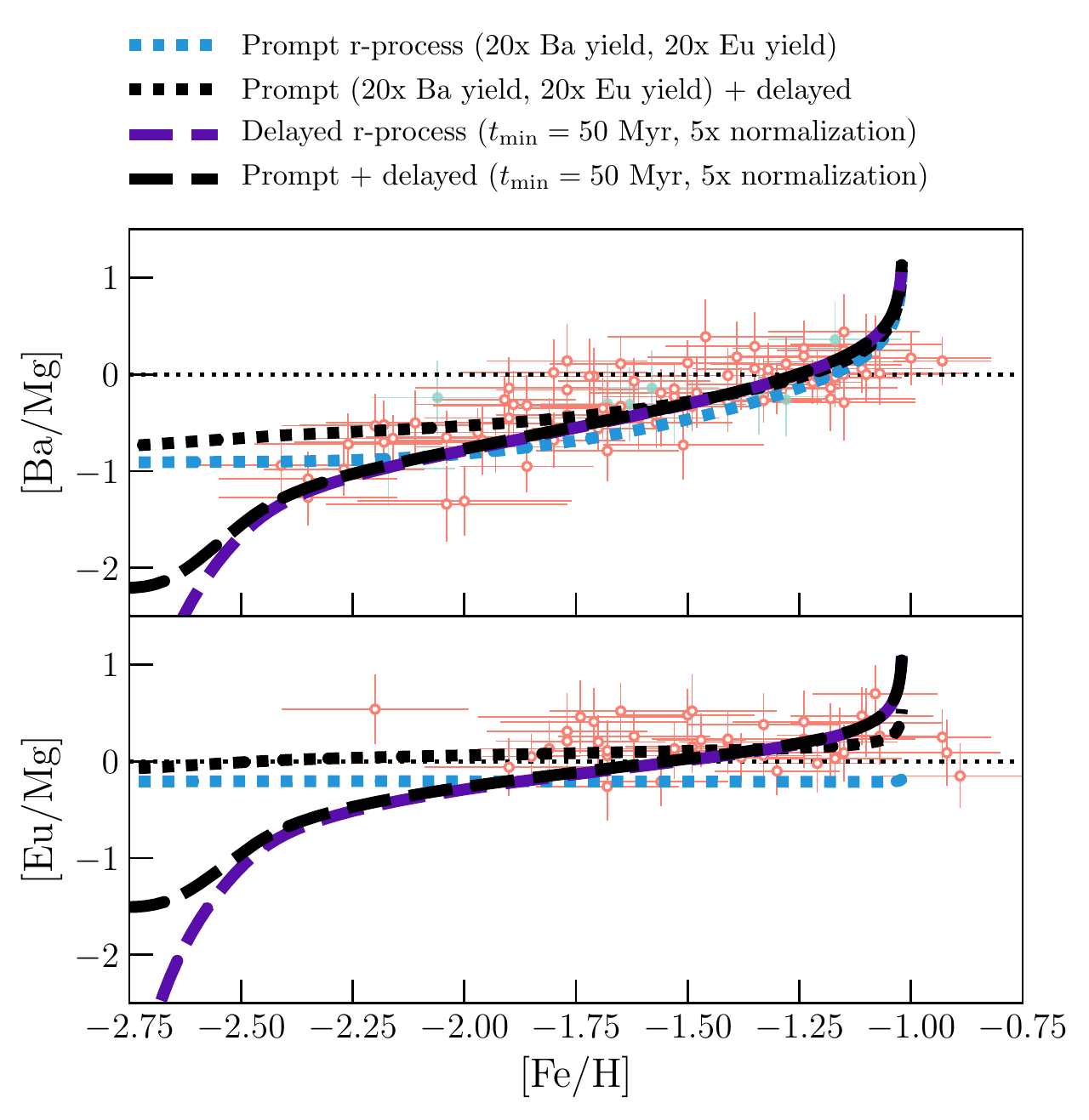}
    \caption{Comparisons among different parameterizations of the $r$-process in the best-fit GCE model (solid lines) for barium (top) and europium (bottom). Left: Best-fit GCE models without any r-process production (cyan), with only prompt CCSN-like r-process events (blue), with only delayed r-process events based on the \citet{Simonetti2019} DTD for neutron star mergers (purple), and with both prompt and delayed r-process events (black). Right: GCE models where the $r$-process channels have been modified to better match the observations. The dotted lines indicate the prompt and prompt+delayed channels when the prompt $r$-process yields have been enhanced, while the dashed lines indicate the delayed and prompt+delayed channels when the DTD has been modified. Points denote observations from DEIMOS (filled blue points) and DART (empty orange points).}
    \label{fig:rprocess}
\end{figure*}

\subsection{Probing $r$-process nucleosynthesis using Ba and Eu}
\label{sec:ba}

As shown in Figure~\ref{fig:gceresult_scl}, our simple GCE model is largely able to capture the behavior of the barium abundances produced by the slow neutron-capture process $\mathrm{[Ba/Fe]}_{s}$.
However, as discussed in Section~\ref{sec:observations}, the rapid neutron-capture process ($r$-process) produces a non-negligible amount of barium. It also produces the majority of europium.
Because our initial model does not include any $r$-process contribution, it significantly underpredicts the total Ba and Eu abundances (solid cyan lines in Figure~\ref{fig:rprocess}).

Despite recent work identifying neutron star mergers (NSMs) as a key site of the $r$-process \citep[see, e.g.,][and references therein]{Rosswog2018,Shibata2019}, the details of $r$-process nucleosynthesis are still uncertain.
A full analysis of $r$-process nucleosynthesis is beyond the scope of this work, but our GCE model can place some constraints on $r$-process timescales and yields.
In particular, we can constrain the rough timescale of $r$-process nucleosynthesis: does the $r$-process primarily occur in prompt events (such as high-mass or rapidly-rotating CCSNe), or relatively more delayed events (such as NSMs)?

To test this, we use our best-fit GCE model (parameters from Table~\ref{tab:params}) and modify the input yields to simulate prompt or delayed $r$-process events.
To model prompt $r$-process events, we add a contribution to the CCSN yields for Ba and Eu.
We use Ba yields predicted by \citet{Li2014} (see Appendix~\ref{sec:appendix_yields} for details) and assume a universal $r$-process ratio of $\mathrm{[Ba/Eu]}\sim-0.7$ \citep[][]{Sneden2008} to compute Eu yields.
To model delayed $r$-process events, we assume a relatively steep NSM DTD in the form of a $t^{-1.5}$ power law \citep[e.g.,][]{Cote2019,Simonetti2019}.
Specifically, we initially adopt the following NSM DTD from \citet{Simonetti2019}:
\begin{equation}
    \Psi_{\mathrm{Ia}} = \left\{ \begin{array}{ll}
            0 & t_{\mathrm{delay}} < 0.01~\mathrm{Gyr} \\
            (10^{-4}~\mathrm{Gyr}^{-1}~M_{\odot}^{-1})\left(\frac{t_{\mathrm{delay}}}{\mathrm{Gyr}}\right)^{-1.5} & t_{\mathrm{delay}} \geq 0.01~\mathrm{Gyr}
        \end{array}\right.
    \label{eq:nsm_dtd}
\end{equation}
We assume that each NSM produces the same amount of $r$-process elements: $M_{\mathrm{Ba}}=2.3\times10^{-6}~M_{\odot}$ and $M_{\mathrm{Eu}}=2.3\times10^{-7}~M_{\odot}$, from \citet{Li2014}.

In the left panels of Figure~\ref{fig:rprocess}, we plot the results of this test, showing the model predictions for [Ba/Mg] and [Eu/Mg] as a function of [Fe/H] assuming different $r$-process parameterizations: no $r$-process contributions (solid cyan line), contributions from either prompt or delayed channels (solid blue and purple lines, respectively), and contributions from both channels (solid black line).
We plot [X/Mg] rather than [X/Fe] because Type Ia SNe contribute significantly to Fe at late times and could complicate our interpretation.
Observations from both DEIMOS and DART are plotted for comparison.
A successful model should be consistent with the observed trends: subsolar [Ba/Mg] below $\mathrm{[Fe/H]}\lesssim-1.5$ that increases to near-solar at higher metallicities, and near-solar [Eu/Mg] that is roughly constant as a function of metallicity.

We first consider whether prompt $r$-process events alone can reproduce these observed trends.
Our initial test (left panels of Figure~\ref{fig:rprocess}) finds that the prompt $r$-process channel (solid blue line) appears to significantly underpredict both [Ba/Mg] and [Eu/Mg].
This may suggest that our assumed CCSN yields of Ba and Eu \citep[respectively from][]{Li2014,Cescutti2006} are too low.
We therefore attempt to produce better consistency with observations by increasing the CCSN yields of Ba and Eu, as shown by the dotted lines in the right panels of Figure~\ref{fig:rprocess}.
In order to better match the Ba abundances, the $r$-process yields from typical CCSN-like events must be drastically increased by a factor of 20.
This may correspond physically to some combination of increasing the rate of CCSN-like $r$-process events and increasing the Ba and Eu yields expected from these events.
This enhancement of prompt $r$-process contributions is able to increase the predicted [Ba/Mg] trend at low metallicity, making it consistent with observations at $-2.5\lesssim\mathrm{[Fe/H]}\lesssim -2$ (blue dotted line in the upper right panel of Figure~\ref{fig:rprocess}), but it underpredicts the observed [Ba/Mg] at $-2\lesssim\mathrm{[Fe/H]}\lesssim-1.5$.

Additionally, \citet{Frebel2010} measured abundances in the most metal-poor star in Sculptor and found abundance ratios of $\mathrm{[Fe/H]}\sim-3.8$ and $\mathrm{[Ba/Mg]}\lesssim-1.58$.
This is well below the $\mathrm{[Ba/Mg]}\sim-1$ at low metallicities predicted by the prompt $r$-process channel with enhanced yields.
Since $r$-process events are likely rare \citep[e.g.,][]{Ji2016}, it is possible that this star simply formed before any prompt CCSN-like events could produce barium.
However, the prompt channel alone also underestimates the observed [Eu/Mg] trend (blue dotted line in the lower right panel of Figure~\ref{fig:rprocess}), despite the drastic enhancement in $r$-process yields.
We therefore tentatively conclude that prompt CCSN-like $r$-process events alone may not be able to explain the observed trends, and at least some contribution from a delayed NSM-like process is likely needed.

We now consider the delayed $r$-process channel.
The left panels of Figure~\ref{fig:rprocess} show that the delayed $r$-process channel (solid purple lines) is more consistent with observations than the prompt $r$-process channel (solid blue lines).
The delayed channel also appears to dominate $r$-process nucleosynthesis; prompt $r$-process events do not contribute significantly to the combined prompt+delayed model (solid black lines).
Yet this delayed $r$-process model, based on the \citet{Simonetti2019} NSM DTD (Equation~\ref{eq:nsm_dtd}), underpredicts the observed [Eu/Mg] trend.
It also slightly overpredicts [Ba/Mg] at low [Fe/H] and underpredicts [Ba/Mg] at $-2\lesssim\mathrm{[Fe/H]}\lesssim-1$.

In order to better match observations, we increase the normalization of the delayed $r$-process contributions by a factor of five.
Physically, this can be done either by increasing the normalization of the \citet{Simonetti2019} NSM DTD, or by increasing the Ba and Eu yields expected from individual NSMs. 
Both of these modifications are plausible because both the NSM DTD and the nucleosynthetic yields from NSMs are relatively uncertain.
We also increase the NSM $t_{\mathrm{min}}$ from 10 to 50~Myr.
As shown by the dashed lines in the right panels of Figure~\ref{fig:rprocess}, this allows the delayed $r$-process channel to begin contributing to $r$-process yields at later times than the CCSNe, so that [Ba/Mg] begins increasing at a higher [Fe/H].
This more-delayed NSM DTD produces a steeper [Ba/Mg] trend as a function of metallicity, which is more consistent with the observed trend.
However, it also produces a steeper [Eu/Mg] trend as a function of metallicity, which is less consistent with the observed flat [Eu/Mg] trend \citep[][]{Skuladottir2019}.
We conclude that relatively delayed (with minimum delay times $\gtrsim50$~Myr) $r$-process events alone may not be able to reproduce observed abundance trends in Sculptor, and that a combination of prompt and delayed events is needed.

Our result is consistent with previous results. 
\citet{Duggan2018} found that in multiple Local Group dSphs, the positive trend of [Ba/Fe] as a function of [Fe/H] appears to be steeper than the positive trend of [Mg/Fe] as a function of [Fe/H].
They argued that the primary $r$-process source of Ba must therefore be delayed relative to CCSNe (the primary source of Mg) in order to produce the [Ba/Fe] trend in dwarf galaxies.
In a separate analysis of Sculptor dSph, \citet{Skuladottir2019} use the flat [Eu/Mg] trend to argue the opposite: that the primary source of Eu must \emph{not} be significantly delayed relative to the primary source of Mg.
As our GCE model shows, both prompt and delayed $r$-process channels may be needed to reproduce the observed Ba and Eu trends in Sculptor\@.
This agrees with investigations of the $r$-process in the Milky Way and in the universe \citep[e.g.,][]{Matteucci2014,Cescutti2015,Wehmeyer2015,Cote2019,Siegel2019}, which find that a combination of delayed NSMs and prompt CCSN-like events can successfully reproduce the observed trend and scatter of [Eu/Fe] in the Milky Way.

\section{Conclusions}
\label{sec:conclusion}
We have used a simple one-zone GCE model (Equation~\ref{eq:model}) to simultaneously understand the SFH and chemical evolution history of Sculptor dSph.
This model is able to fit both the metallicity distribution function and the abundance patterns of seven elements.
Like previous one-zone GCE models, our model fits the trends of the $\alpha$ elements Mg, Si, and Ca, which probe CCSNe (delay times of $\sim10$~Myr after star formation), and Fe, which is predominantly produced by Type Ia supernovae (delay times of $\gtrsim100$~Myr).
Our model is also able to fit the observed abundances of C and Ba, which trace nucleosynthesis in AGB stars (delay times of $\gtrsim10$~Myr after star formation).

Our best-fit model (Figure~\ref{fig:gceresult_scl}) indicates that Sculptor dSph had an SFH with a total star formation duration of $\sim0.92$~Gyr.
As shown in Figure~\ref{fig:sfh_scl}, this is in contrast with some photometric measurements \citep[e.g.,][]{deBoer2012,Savino2018}, which found that Sculptor dSph has an extended SFH spanning $6-7$~Gyr, but qualitatively consistent with other recent estimates from deep photometry that find a relatively short and ancient SFH \citep[e.g.,][]{Weisz2014,Bettinelli2019}.
It is also quantitatively consistent with other SFHs derived from GCE models \citep[][]{Kirby2011, Vincenzo2016}, which predict SFH durations of $1-2$~Gyr.
However, the star formation duration of $\sim0.92$~Gyr predicted by our GCE model is shorter than the estimates from even the deepest photometric measurements, which found that the majority of star formation ended after $\sim3-4$~Gyr.
This discrepancy may be partly due to various assumptions of our model---in particular, we find that changing the stellar IMF or the Type Ia DTD may increase the duration of the SFH by up to 50\% (from $\sim0.92$ to $\sim1.4$~Gyr).
Alternatively, our model's short SFH estimate may be a real result; although spectroscopic measurements cannot determine absolute ages, they may be more sensitive to age spreads in old stellar populations than photometric measurements.

Not only is our model able to probe Sculptor dSph's SFH, it can also provide insight into the nucleosynthesis of individual elements in the galaxy.
In our model, we parameterized the nucleosynthetic yields from CCSNe and AGB stars using analytic functions.
For the elements most sensitive to variations in the nucleosynthetic yields, we defined free parameters to vary the yields.
The best-fit free parameters produce yield patterns that resemble those from the \citet{Nomoto2013} CCSN yields and the Stromlo \citep[][]{Karakas2016,Karakas2018} AGB yields.

We also used our GCE model to test the nucleosynthesis of elements that were not used to fit the model.
Mn and Ni are iron-peak elements that are sensitive to the physics of Type Ia supernovae.
We found that the observed [Mn/Fe] and [Ni/Fe] trends are best reproduced by the sub-$M_{\mathrm{Ch}}$ ($1.1~M_{\odot}$) Type Ia supernova model from \citet{Leung2020}.
Similarly, Ba and Eu are elements that can be used to trace $r$-process nucleosynthesis.
We find that a combination of prompt CCSN-like events and delayed $r$-process events (with minimum delay times $\gtrsim50$~Myr) are required to reproduce the observed trends of [Ba/Mg] and [Eu/Mg].

We have shown that a simple GCE model can be used to probe both the star formation and chemical evolution of Sculptor dSph by fitting to the observed abundances of elements produced by multiple kinds of nucleosynthetic events.
This method could easily be applied to other Local Group dSphs with available stellar spectroscopy.
This would not only provide a complementary approach to photometrically derived SFHs but could also be used to test nucleosynthetic processes in different environments---for example, previous measurements of iron-peak elements have suggested that the dominant channel of Type Ia supernovae might depend on a galaxy's SFH \citep{Kirby2019,delosReyes2020}.
Finally, as always, this work could be extended with a larger sample size of observations, more precise spectroscopic measurements, or a more sophisticated model.

\acknowledgments
{This material is based upon work supported by the National Science Foundation under grant No.\ AST-1847909. 
MAdlR and EHN acknowledge the financial support of the NSF Graduate Research Fellowship Program.
ENK gratefully acknowledges support from a Cottrell Scholar award administered by the Research Corporation for Science Advancement.
APJ acknowledges support from a Carnegie Fellowship and the Thacher Research Award in Astronomy.

There are many communities without whom this work would not have been possible. 
We acknowledge that this work is rooted in Western scientific practices and is the material product of a long and complex history of settler-colonialism. MAdlR, ENK, EHN, and APJ wish to recognize their status as settlers on the traditional and unceded territory of the Tongva peoples, and to recognize that the astronomical observations described in this paper were only possible because of the dispossession of Maunakea from K$\bar{\mathrm{a}}$naka Maoli. We hope to work toward a scientific practice guided by pono and a future in which we all honor the land.

Finally, we would like to express our deep gratitude to the staff at academic and telescope facilities, particularly those whose communities are excluded from the academic system but whose labor maintains spaces for scientific inquiry.}

\vspace{5mm}
\facilities{Keck:II (DEIMOS)}

\software{
Matplotlib \citep{matplotlib}, 
Astropy \citep{astropy},
Scipy \citep{scipy},
emcee \citep{Foreman-Mackey2013}
}

\appendix
\section{Nucleosynthetic yield parameterizations}
\label{sec:appendix_yields}
As described in Section~\ref{sec:yields}, rather than select particular model yield sets, we represent the nucleosynthetic yields from supernovae and AGB stars with analytic functions.
Parameters of these functions are varied in conjunction with the other model parameters.

\begin{figure}[h!]
    \centering
    \epsscale{1.1}
    \plotone{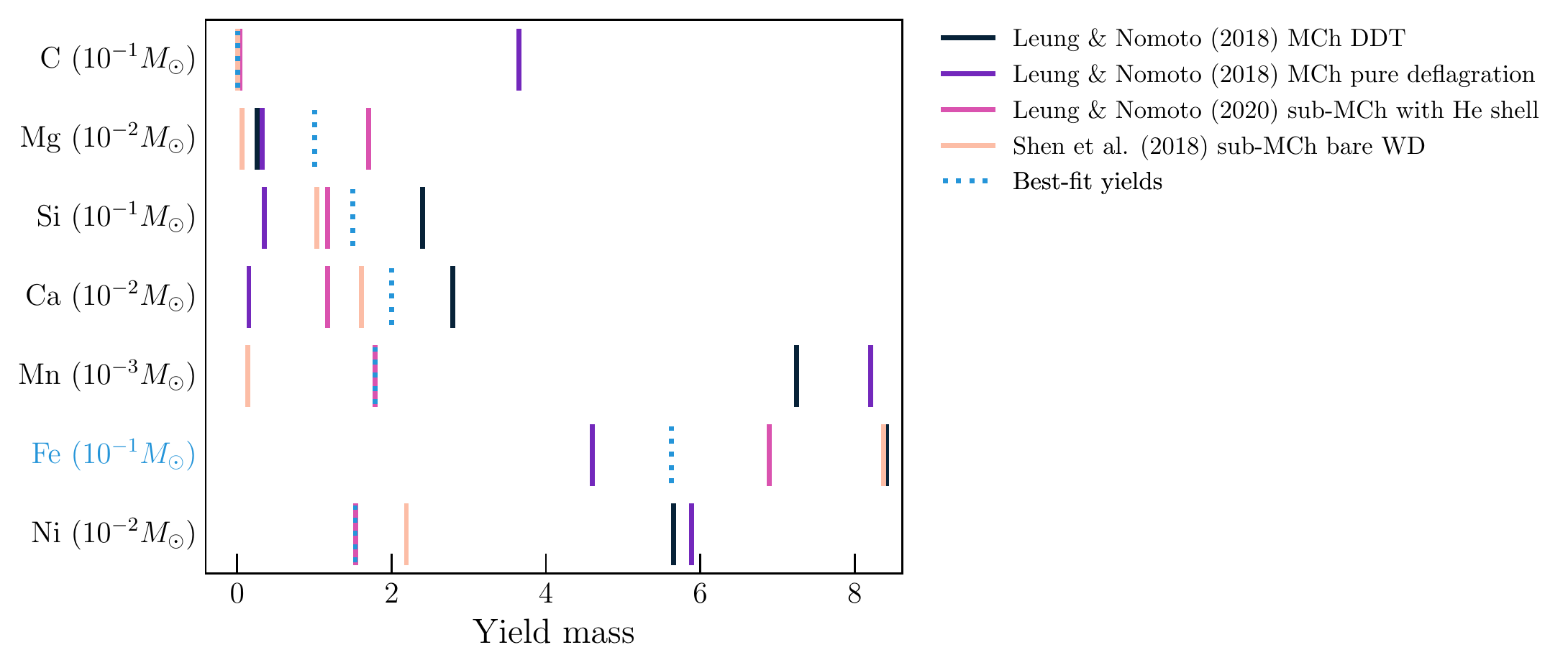}
    \caption{Nucleosynthetic yields from Type Ia supernovae. Colors denote different models. Yields are assumed to be independent of metallicity. Dotted blue lines show the parameterized yields, qualitatively chosen as described in Appendix~\ref{sec:appendix_yields}. A free parameter is used to fit the yield of Fe (labeled with blue text), so the dotted blue line for Fe is the best-fit value from the fiducial GCE model (Table~\ref{tab:params}).}
    \label{fig:iasnyields}
\end{figure}

The yields from Type Ia SNe are assumed to be independent of progenitor mass and metallicity. 
We plot the Type Ia SN yields from different models in Figure~\ref{fig:iasnyields}, as well as the yields chosen for our model (dotted blue lines).
For C, Mg, Si, and Ca, the Type Ia SN yields were selected to be within the range of model yields, although the exact values do not matter because these yields do not significantly impact the final abundance trends.
Manganese and nickel are not included in the GCE model fitting, so for our initial model we arbitrarily choose the yields from \citet{Leung2020} for these elements (see Section~\ref{sec:mnni}).
The yield of Fe from Type Ia SNe is a free parameter ($\mathrm{Fe}_{\mathrm{Ia}}$); Figure~\ref{fig:iasnyields} shows the best-fit value from our fiducial model (Table~\ref{tab:params}).

\begin{figure}[t!]
    \centering
    \epsscale{1.1}
    \plotone{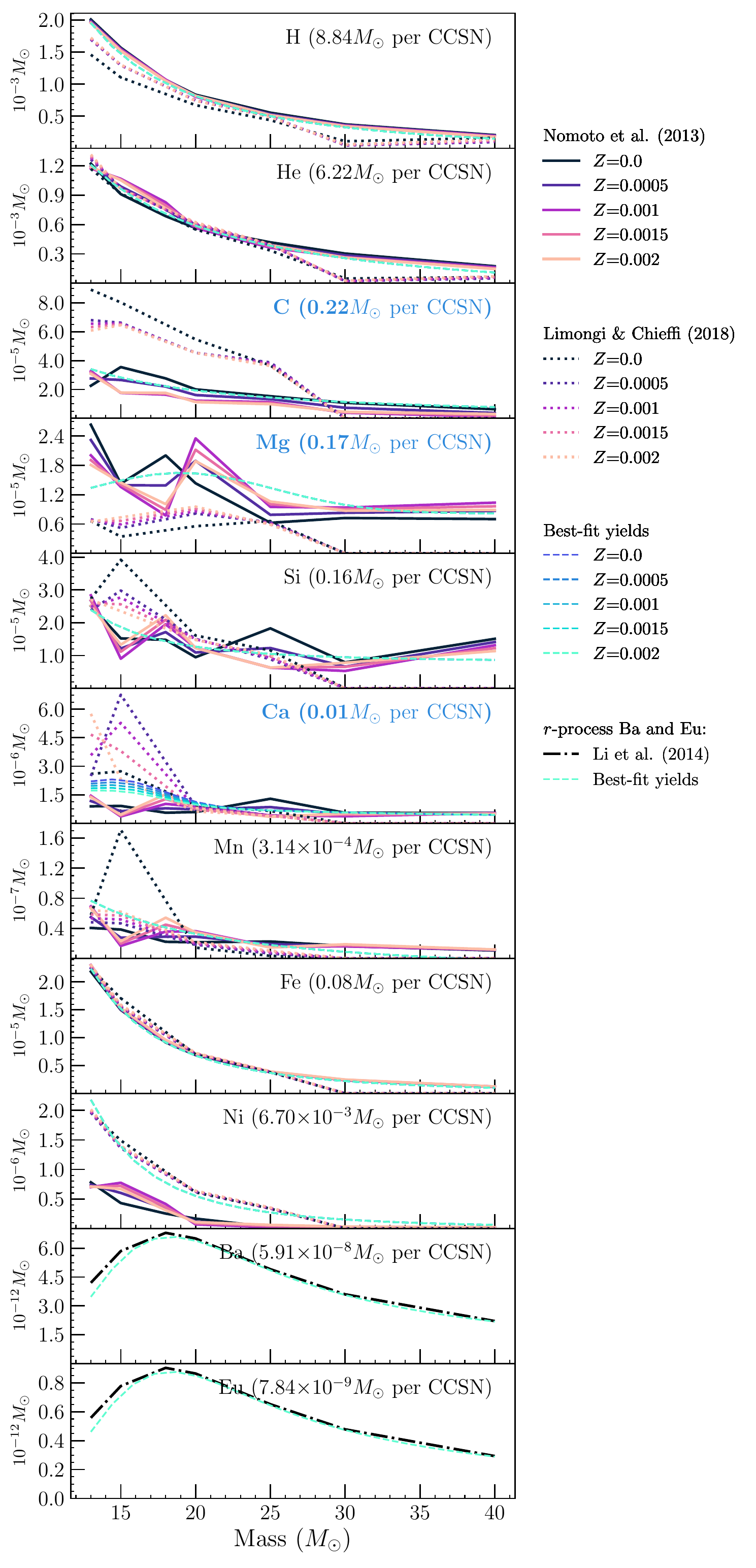}
    \caption{Nucleosynthetic yields from core-collapse supernovae as a function of initial stellar mass. Colors denote different metallicities; yields have been linearly interpolated to match the same metallicity range. Overplotted dashed lines show parameterized yields for illustration. For elements labeled with bold blue text, free parameters are used to fit the GCE model. As described in the text, these yields are weighted by a \citet{Kroupa1993} IMF. To aid in interpretation, we list the average $Z=0.002$ yield per CCSN---i.e., the $Z=0.002$ best-fit analytic function integrated over the yield sets' CCSN progenitor mass range $13-40~M_{\odot}$ and multiplied by a factor of 500 (since roughly 1 CCSN is produced for every $500~M_{\odot}$ of stars formed)---in the upper right corner of each plot.}
    \label{fig:ccsnyields}
\end{figure}

\begin{figure}[t!]
    \centering
    \epsscale{1.18}
    \plotone{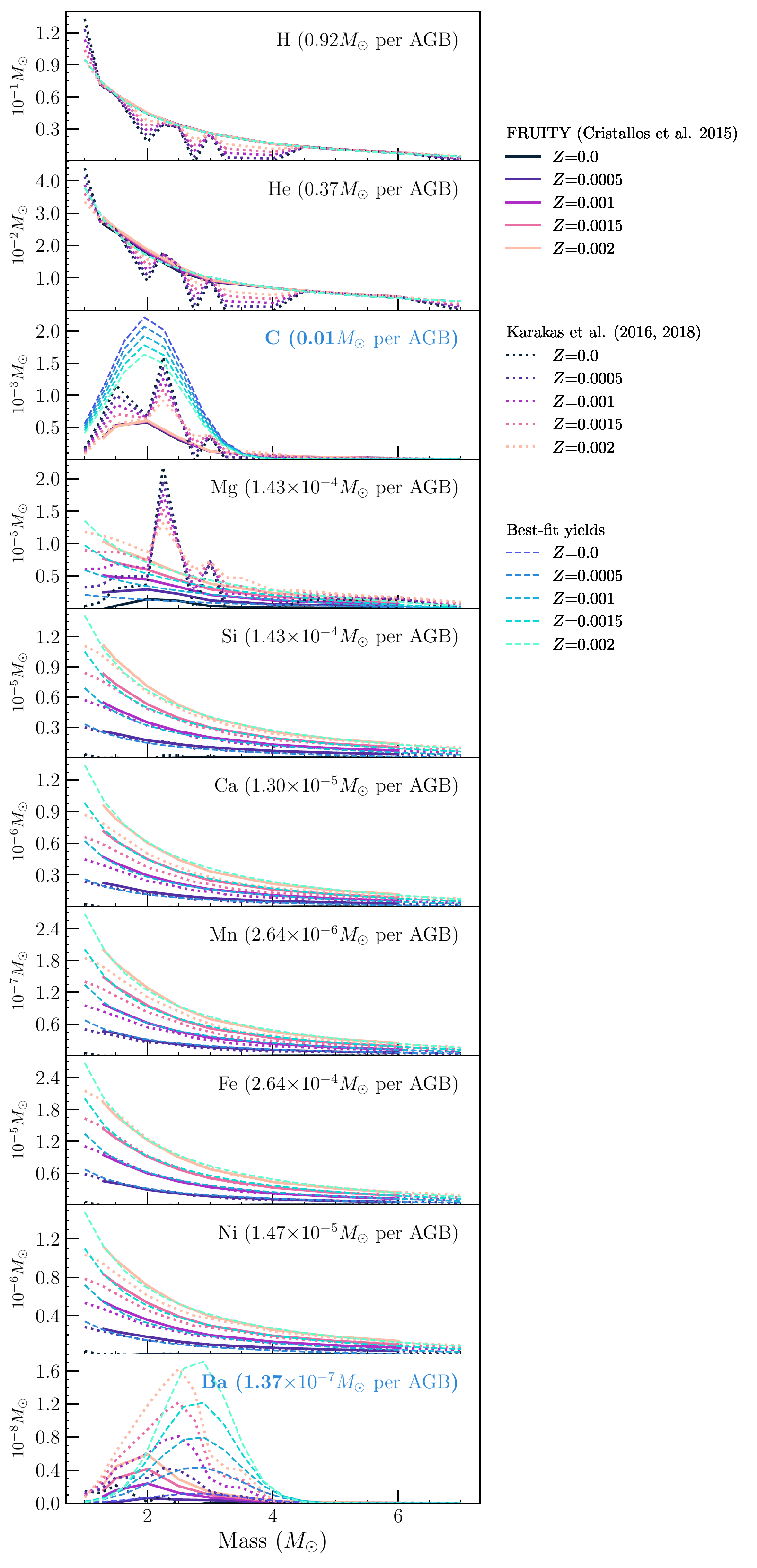}
    \caption{Similar to Figure~\ref{fig:ccsnyields}, but plotting nucleosynthetic yields from AGB stars. To compute the average $Z=0.002$ yield per AGB star, we integrate over the yield sets' AGB progenitor mass range $1-7~M_{\odot}$ and multiply by a factor of 5.7 (since roughly 1 AGB star is produced for every $5.7~M_{\odot}$ of stars formed).}
    \label{fig:agbyields}
\end{figure}

For CCSN and AGB yields, we fit the model yield sets (Table~\ref{tab:yields}) with combinations of exponential and Gaussian functions.
We chose functions that qualitatively approximate the shapes of the theoretical yields as functions of progenitor mass ($M$) and metallicity ($Z$).
In particular, we fit yields that are mass-weighted by a \citet{Kroupa1993} IMF (i.e., multiplied by the IMF $dN/dM$).
The analytic functions can then be integrated over a range of progenitor stellar masses to compute the total yield of a given element that will be produced by stars in that mass range after an instantaneous $1~M_{\odot}$ star formation burst.
These functions are listed in Table~\ref{tab:yieldfuncs} and illustrated as dotted blue lines in Figures~\ref{fig:ccsnyields} and \ref{fig:agbyields} for CCSNe and AGB stars, respectively.
The theoretical yield sets are also plotted for comparison (solid and dotted lines).
For completeness, Figure~\ref{fig:ccsnyields} also illustrates the \citet{Li2014} $r$-process yields from CCSNe assumed in Section~\ref{sec:ba}.

\begin{deluxetable*}{lll}[t!]
\tablecolumns{3} 
\tablecaption{Analytic functions describing IMF-weighted CCSN and AGB yields. \label{tab:yieldfuncs}} 
\rotate
\tablehead{ 
\colhead{Element} & \colhead{CCSN yield ($M_{\odot}$)} & \colhead{AGB yield ($M_{\odot}$)}
}
\startdata
H & $10^{-3}\left(255\left(\frac{M}{M_{\odot}}\right)^{-1.88} - 0.1\right)$ & $10^{-1} \left(1.1\left(\frac{M}{M_{\odot}}\right)^{-0.9} - 0.15\right)$ \\
He & $10^{-3}\left(45\left(\frac{M}{M_{\odot}}\right)^{-1.35} - 0.2\right)$ & $10^{-2}\left(4\left(\frac{M}{M_{\odot}}\right)^{-1.07} - 0.22\right)$\\
C & $10^{-5}\left(100\left(\frac{M}{M_{\odot}}\right)^{-\mathrm{expC}_{\mathrm{II}}}\right)$ & $10^{-3}\left(\mathrm{normC}_{\mathrm{AGB}}\left(1.68-220\frac{Z}{Z_{\odot}}\right)\mathcal{G}\left(\frac{M}{M_{\odot}},2,0.6\right)\right)$\\
Mg & $10^{-5}\left(\mathrm{normMg}_{\mathrm{II}} + 13\mathcal{G}\left(\frac{M}{M_{\odot}},19,6.24\right)\right)$ & $10^{-5}\left(\left(400\frac{Z}{Z_{\odot}} + 1.1\right)\left(\frac{M}{M_{\odot}}\right)^{0.08 - 340(Z/Z_{\odot})} + \left(360\frac{Z}{Z_{\odot}} - 1.27\right)\right)$\\
Si & $10^{-5}\left(2260\left(\frac{M}{M_{\odot}}\right)^{-2.83} + 0.8 \right)$ & $10^{-5}\left(\left(800\frac{Z}{Z_{\odot}}\right)\left(\frac{M}{M_{\odot}}\right)^{-0.9} - \left(80\frac{Z}{Z_{\odot}} + 0.03\right)\right)$ \\
Ca & $10^{-6}\left(15.4\left(\frac{M}{M_{\odot}}\right)^{-1} + \mathrm{normCa}_{\mathrm{II}} \left(40-10^{4}\frac{Z}{Z_{\odot}}\right)\mathcal{G}\left(\frac{M}{M_{\odot}},15,3\right) + 0.06 \right)$ & $10^{-6}\left(\left(800\frac{Z}{Z_{\odot}}-0.1\right)\left(\frac{M}{M_{\odot}}\right)^{-0.96} - 80\frac{Z}{Z_{\odot}}\right)$ \\
Mn & $10^{-7}\left(30\left(\frac{M}{M_{\odot}}\right)^{-1.32} - 0.25\right)$ & $10^{-7}\left(1500\frac{Z}{Z_{\odot}} \left(\frac{M}{M_{\odot}}\right)^{-0.95} - 160\frac{Z}{Z_{\odot}}\right)$ \\
Fe & $10^{-5}\left(2722 \left(\frac{M}{M_{\odot}}\right)^{-2.77}\right)$ & $10^{-5}\left(1500\frac{Z}{Z_{\odot}} \left(\frac{M}{M_{\odot}}\right)^{-0.95} - 160\frac{Z}{Z_{\odot}}\right)$ \\
Ni$^{a}$ & $10^{-7}\left(8000\left(\frac{M}{M_{\odot}}\right)^{-3.2}\right)$ & $10^{-6}\left(840\frac{Z}{Z_{\odot}} \left(\frac{M}{M_{\odot}}\right)^{-0.92} - \left(80\frac{Z}{Z_{\odot}} + 0.04\right)\right)$\\
Ba$^{b}$ & $10^{-12}\left(1560\left(\frac{M}{M_{\odot}}\right)^{-1.80} + 0.14 - 480\mathcal{G}\left(\frac{M}{M_{\odot}},5,5.5\right)\right)$ & $10^{-8}\left(\mathrm{normBa}_{\mathrm{AGB}} \left(10^3\frac{Z}{Z_{\odot}} + 0.2\right)\mathcal{G}\left(\frac{M}{M_{\odot}},\mathrm{meanBa}_{\mathrm{AGB}},0.75-100\frac{Z}{Z_{\odot}}\right)\right)$ \\
Eu$^{b}$ & $\frac{1}{7.53}10^{-12}\left(1560\left(\frac{M}{M_{\odot}}\right)^{-1.80} + 0.14 - 480\mathcal{G}\left(\frac{M}{M_{\odot}},5,5.5\right)\right)$ & $10^{-11}\left(\left(3400\frac{Z}{Z_{\odot}} + 0.4\right)\mathcal{G}\left(\frac{M}{M_{\odot}},2,0.65\right)\right)$\\
\enddata
\tablenotetext{a}{The function for CCSN yields of Ni was chosen to fit the yields from \citet{Limongi2018}. Since Ni is not included in fitting our GCE model (Section~\ref{sec:gce}), this does not affect any of our results; however, we find that the \citet{Nomoto2013} yields significantly underpredict the observed [Ni/Fe] at low [Fe/H].}
\tablenotetext{b}{We assume that the CCSN Ba and Eu yields are produced by the $r$-process. Our fiducial model assumes that zero Ba and Eu are produced by CCSNe. The functions describing CCSN yields listed here are for the $r$-process; as described in the text, we use the Ba yields from \citet{Li2014} and scale by the universal $r$-process ratio $\mathrm{[Ba/Eu]\sim-0.7}$ \citep[see][]{Sneden2008} to obtain the Eu yields.}
\tablecomments{As described in the text, these yields are weighted by a \citet{Kroupa1993} IMF. In these equations, we use the notation $\mathcal{G}\left(\frac{M}{M_{\odot}},\mu,\sigma\right)$ to denote a normalized Gaussian function as a function of mass: $\mathcal{G}\left(\frac{M}{M_{\odot}},\mu,\sigma\right) = \frac{1}{\sigma\sqrt{2\pi}}\exp {\left(-{\frac {((M/M_{\odot})-\mu)^{2}}{2\sigma^{2}}}\right)}$}
\end{deluxetable*}

\bibliographystyle{aasjournal}
\bibliography{references}

\end{document}